\documentclass[10pt,journal,compsoc]{IEEEtran}
\usepackage[sort&compress, numbers]{natbib}
\usepackage{verbatim}
\usepackage{caption}
\usepackage{subcaption}
\usepackage{booktabs}
\usepackage{relsize}
\usepackage{colortbl}
\usepackage{enumerate}
\usepackage{amsmath}
\usepackage{tikz}
\usepackage{pgf}
\usepackage{pbox}
\usepackage{rotating}
\usepackage{multirow}
\usepackage{balance}
\usepackage{tablefootnote}
\usepackage[bookmarks=false]{hyperref}
\usepackage{flushend}
\usepackage{extarrows}
\usepackage{cleveref}[2012/02/15]
\usepackage{arydshln}
\crefformat{footnote}{#2\footnotemark[#1]#3}
\usepackage{soul}

\newcommand{\startlist}{\begin{list}{\labelitemi}{\leftmargin=1em}\setlength{\itemsep}{-1mm}}
	\newcommand{\stoplist}{\end{list}}

\newcommand{\smallsection}[1]{\noindent {\bf \underline{#1}}.\hspace{1mm}}
\newcommand{\ea}{{\em et al.}}

\newcommand{\revised}[2]{#2}
\newcommand{\revisedTextOnly}[1]{#1}

\newcommand{\rqi}{How do practitioners perceive SQA planning activities?}

\newcommand{\rqii}{How do practitioners perceive our proposed four types of guidance to support SQA planning?}

\newcommand{\rqiii}{How effective are the rule-based explanations generated by our SQAPlanner approach when compared to the state-of-the-art approaches?}

\newcommand{\rqiv}{How stable are the rule-based explanations generated by our SQAPlanner approach when they are regenerated?}

\newcommand{\rqv}{How applicable are the rule-based explanations generated by our SQAPlanner approach to minimize the risk of having defects in the subsequent releases?}

\newcommand{\gli}{What  are  risky  practices  that  lead  a  model  to predict a file as defective?}

\newcommand{\glii}{What are the non-risky practices that lead a model to predict a file as clean?}

\newcommand{\gliii}{What are practices to avoid to not increase the risk of having defects?}

\newcommand{\gliv}{What are the practices to follow to decrease the risk of having defects?}

\newcommand{\rqvi}{How do practitioners perceive the visualization of SQAPlanner when comparing to the visualization of the state-of-the-art?}

\newcommand{\rqvii}{How do practitioners perceive the actual guidance generated by our SQAPlanner?}

\usepackage{mathrsfs}
\usepackage[linesnumbered,ruled,vlined]{algorithm2e}
\usepackage{graphicx}
\usepackage{csquotes}
\usepackage[T1]{fontenc}
\usepackage{multirow}
\usepackage{makecell, caption}
\usepackage{tabularx}

\newlength\mylen
\newcommand\myinput[1]{%
    \settowidth\mylen{\KwIn{}}%
    \setlength\hangindent{\mylen}%
    \hspace*{\mylen}#1\\}

\newcommand{\LineNumbered}{%
    \setboolean{algocf@linesnumbered}{true}%
    \renewcommand{\algocf@linesnumbered}{\everypar={\nl}}}%

\let\oldnl\nl
\newcommand{\nonl}{\renewcommand{\nl}{\let\nl\oldnl}}

\newcolumntype{L}{>{\raggedright\arraybackslash}X}

\graphicspath{{figures/}}

\usepackage[noend]{algpseudocode}
\usepackage{amsmath}
\usepackage{mathtools}
\usepackage{xparse}

\usepackage{mathrsfs}

\DeclareDocumentCommand{\outAlgo}{ m O{\gets} }{%
	{\rlap{$#1$} \hspace{0.7cm}\hphantom{text}$#2$}%
}

\DeclareDocumentCommand{\jingAlgo}{ m O{-} }{%
	{\rlap{$#1$} \hspace{0.7cm}\hphantom{text}$#2$}%
}

\begin{document}

\title{SQAPlanner: Generating Data-Informed Software Quality Improvement Plans}

\author{Dilini Rajapaksha,
        Chakkrit Tantithamthavorn, Jirayus Jiarpakdee, \\
        Christoph Bergmeir, John Grundy, and Wray Buntine
\IEEEcompsocitemizethanks{
\IEEEcompsocthanksitem D. Rajapaksha, C. Tantithamthavorn, J. Jiarpakdee C. Bergmeir, J. Grundy, and W. Buntine are with the Faculty of Information Technology, Monash University, Melbourne, Australia.\protect\\
E-mail: \{dilini.rajapakshahewaranasinghage, chakkrit, jirayus.jiarpakdee, christoph.bergmeir,  john.grundy, wray.buntine\}@monash.edu
\IEEEcompsocthanksitem Corresponding author: C. Tantithamthavorn.
}
}


\IEEEtitleabstractindextext{%

\begin{abstract}
\revisedTextOnly{
Software Quality Assurance (SQA) planning aims to define proactive plans, such as defining maximum file size, to prevent the occurrence of software defects in future releases.
To aid this, \emph{defect prediction models} have been proposed to generate insights as the most important factors that are associated with software quality.
Such insights that are derived from traditional defect models are far from actionable---i.e., practitioners still do not know what they should do or avoid to decrease the risk of having defects, and what is the risk threshold for each metric.
A lack of actionable guidance and risk threshold can lead to inefficient and ineffective SQA planning processes. 
In this paper, we investigate the practitioners' perceptions of current SQA planning activities, current challenges of such SQA planning activities, and propose four types of guidance to support SQA planning.
We then propose and evaluate our AI-Driven SQAPlanner approach, a novel approach for generating four types of guidance and their associated risk thresholds in the form of rule-based explanations for the predictions of defect prediction models.
Finally, we develop and evaluate an information visualization for our SQAPlanner approach.
Through the use of qualitative survey and empirical evaluation, our results lead us to conclude that SQAPlanner is needed, effective, stable, and practically applicable.
We also find that 80\% of our survey respondents perceived that our visualization is more actionable.
Thus, our SQAPlanner paves a way for novel research in actionable software analytics---i.e., generating actionable guidance on what should practitioners do and not do to decrease the risk of having defects to support SQA planning.}
\end{abstract}

\begin{IEEEkeywords}
Software Quality Assurance, SQA Planning, Actionable Software Analytics, Explainable AI.
\end{IEEEkeywords}}

\maketitle
\IEEEdisplaynontitleabstractindextext
\IEEEpeerreviewmaketitle

\section{Introduction}

\revisedTextOnly{
Software Quality Assurance (SQA) planning is the process of developing proactive SQA plans.
One of the most important SQA activities is to define development policies and their associated risk thresholds~\cite{galin2018software} (e.g., defining the maximum file size, the maximum code complexity, and the minimum degree of code ownership).
Such SQA plans will be later enforced for a whole team to ensure the highest quality of software systems.
These policies are essential to improve software quality and software maintainability~\cite{maxim2016introduction}.

\revised{R2.15}{
Recently, top software companies have released several commercial AI-driven defect prediction tools.}
For example, Microsoft's Code Defect AI, Amazon's CodeGuru.
Such tools heavily rely on the concept of defect prediction models that have been well-studied in the past decades~\cite{hall2012systematic}.
In particular, Microsoft's Code Defect AI is built on top of the concept of explainable Just-In-Time defect prediction~\cite{jiarpakdee2020xai4se, tantithamthavorn2020explainable}---i.e., explaining the predictions of defect models using a LIME model-agnostic technique~\cite{ribeiro2016should}.
The crux of Microsoft's Code Defect AI tool is similar to the recent parallel work by Jiarpakdee~\ea~\cite{jiarpakdee2020xai4se} who also suggested to use a LIME model-agnostic technique to explain the predictions of defect models.

However, these current state-of-the-art defect prediction approaches can only indicate the most important features, which are still far from actionable. 
Thus, practitioners still do not know (1) what they should do to decrease the risk of having defects, and what they should avoid to not increase the risk of having defects and (2) what is a risk threshold for each metric (e.g., how large is a file size that would be risky? and how small is a file size that would be non-risky?). 

A lack of actionable guidance and its risk threshold can lead to inefficient and ineffective SQA planning processes. 
Such ineffective SQA planning processes will result in the recurrence of software defects, slow project progress, high costs of development, unsatisfactory software products, and unhappy end-users. 
These challenges are very significant to the practical applications of defect prediction models, but still remain largely unexplored.

We aim to help practitioners to make better data-informed SQA planning decisions by generating actionable guidance derived from defect prediction models.
Thus, we first propose the following four types of guidance to support SQA planning:

\revised{R1.3}{
\begin{enumerate}[{\bf (G1)}]
\item \textbf{Risky current practices that lead the defect model to predict a file as defective} are needed to help practitioners understand what are the current risky practices. 
\item \textbf{Non-risky current practices that lead the defect model to predict a file as clean} are needed to help practitioners understand what are the non-risky current practices. 
\item \textbf{Potential practices to avoid to not increase the risk of having defects} are needed to help practitioners understand which currently not implemented practices to avoid to not increase the risk of having defects. 
\item \textbf{Potential practices to follow to decrease the risk of having defects} are needed to help practitioners understand which practices to newly implement to decrease the risk of having defects.
\end{enumerate}}

To achieve this aim, our research study has the following 3 key objectives: 

\begin{enumerate}[{\bf (Obj1)}]
    \item Investigating practitioners' perceptions and challenges of carrying out current SQA planning activities and the perceptions of our proposed four types of guidance;
    \item Developing and evaluating our novel SQAPlanner approach and comparing it with state-of-the-art approaches;
    \item Developing and evaluating an information visualization for our SQAPlanner approach and comparing it with the visualization of Microsoft’s Code Defect AI tool.
\end{enumerate}

To achieve the first objective, we first conducted a qualitative survey with practitioners to address the following research questions:

\begin{enumerate}[{\bf (RQ1)}]
\item {\bf \rqi}
For SQA planning activities, 86\% of the respondents perceived as important and 70\% perceived as being used in practice. However, 66\% perceived as time-consuming and 58\% perceived as difficult, indicating that a data-informed SQA planning tool is needed to support QA teams for better data-informed decision-making and policy-making.
\item {\bf \rqii}
Both (G1) the guidance on risky current practices that lead a model to predict a file as defective and (G4) the guidance on the potential practices to follow to decrease the risk of having defects are perceived as among the most useful, most important, and most considered willingness to adopt by the respondents.
\end{enumerate}

Motivated by the findings of RQ1 and RQ2, we proposed an AI-Driven SQAPlanner---i.e., an approach to generate four types of guidance in the form of rule-based explanations~\cite{rajapaksha2019lormika} to support data-informed SQA planning.
\emph{Our AI-Driven SQAPlanner is a significant advancement over the LIME model-agnostic technique~\cite{ribeiro2016should}}, since LIME only indicates what factors are the most important to support the predictions towards defective (G1) and clean (G2) classes, while our AI-Driven SQAPlanner can  additionally provide actionable guidance on what should developers avoid (G3) and should do (G4) to decrease the risk of having defects.
Then, we conduct an empirical evaluation to evaluate our SQAPlanner approach and compare with two state-of-the-art local rule-based model-agnostic techniques (i.e., Anchor~\cite{ribeiro2018anchors} (i.e., an extension of LIME~\cite{ribeiro2016should}), LORE~\cite{guidotti2018local}).
Through a case study of 32 releases across 9 open-source software projects, we addressed the following research questions:

\begin{enumerate}[{\bf (RQ3)}]
\item {\bf \rqiii}
The rule-based guidance generated by our SQAPlanner approach achieves the highest coverage (at the median 89\%), confidence (at the median 99\%), and lift scores (at the median 6.6) when comparing to baseline techniques.

\item[{\bf (RQ4)}] {\bf \rqiv}
Our SQAPlanner approach produces the most consistent (a median Jaccard coefficient of 0.92) rule-based guidance when comparing to baseline techniques, suggesting that our approach can generate the most stable rule-based guidance when they are regenerated.

\item[{\bf (RQ5)}] {\bf \rqv}
For 55\%-87\% of the defective files, our SQAPlanner approach can generate rule-based guidance that is applicable to the subsequent release to decrease the risk of having defects.

\end{enumerate}

To evaluate the practical usefulness of our SQAPlanner, we developed a proof-of-concept prototype to visualize the actual generated actionable guidance. 
The visualization of our SQAPlanner is designed to provide the following key information: (1) the list of guidance that practitioners should follow and should avoid; (2) the actual feature value of that file; and (3) its threshold and range values for practitioners to follow to mitigate the risk of having defects.
Then, we compare our visualization with the visualization of Microsoft's Code Defect AI (see Figure~\ref{Fig:codedefectaiplot}).
Finally, we conducted a qualitative survey to address the following research questions:

\begin{enumerate}[{\bf (RQ6)}]
\item {\bf \rqvi}
80\% of the respondents agree that the visualization of our SQAPlanner is best to provide actionable guidance compared to the visualization of Microsoft's Code Defect AI.
\item[{\bf (RQ7)}] {\bf \rqvii}
63\%-90\% of the respondents agree with the seven statements derived from the actual guidance generated by our SQAPlanner.
\end{enumerate}

The key contributions of this paper are:

\begin{itemize}
\item An empirical investigation of the practitioners' perceptions and their challenges of current SQA planning activities.
\item An empirical investigation of the practitioners' perceptions of our proposed four types of guidance.
\item The development of our novel AI-Driven SQAPlanner approach to generate the proposed four types of guidance in the form of rule-based explanations to better support SQA planning. \revised{R1.2}{The implementation is available at \url{https://github.com/awsm-research/SQAPlanner-implementation}}.
\item The empirical investigation of the effectiveness, the stability, and the applicability of rule-based explanations generated by our SQAPlanner.
\item The development of the visualization of our SQAPlanner approach and the empirical investigation of the practitioners' perceptions on our visualization and the actual guidance.
\end{itemize}
}

The rest of the paper is organized as follows. 
Section~\ref{sec:background} discusses the significance of SQA planning, the limitations of current AI-driven defect prediction tools, and the motivation of the proposed four types of guidance to support SQA planning.
Section~\ref{sec:rqs} presents the overview of our case study and the motivation of the research questions. 
Section~\ref{sec:survey} presents the results of the practitioners' perceptions of SQA planning activities and the four types of guidance to support SQA planning.
Section~\ref{sec:methodology} presents our SQAPlanner approach, while Section~\ref{sec:results} presents the empirical results of our SQAPlanner approach.
Section~\ref{sec:postvalidation} presents the empirical investigation of the visualization of our SQAPlanner and the actual guidance generated by our SQAPlanner approach.
Section~\ref{sec:threats} summarizes the threats to the validity of our study, and Section~\ref{sec:relatedwork} discusses related work.
Finally, Section~\ref{sec:conclusions} draws the conclusions.

\section{Background and Motivation} 
\label{sec:background}

\revisedTextOnly{In this section, we first discuss the significance of Software Quality Assurance (SQA) planning. 
Then, we discuss the limitations of current AI-driven defect prediction tools.
Finally, we propose the four types of guidance to support SQA planning.}

\subsection{``Prevention is better than cure''}

This is a classic principle that is commonly applied to SQA processes to prevent software defects~\cite{kumaresh2010defect}.
It is widely known that the cost of software defects rises significantly if they are discovered later in the process.
Thus, finding and fixing software defects prior to releasing software is usually much cheaper and faster than fixing after the software is released~\cite{boehm2005software}.
Therefore, SQA teams play a critical role in software companies as a gatekeeper, i.e., not allowing software defects to pass through to end-users.

\begin{figure}[t]
    \centering
    \includegraphics[width=\columnwidth]{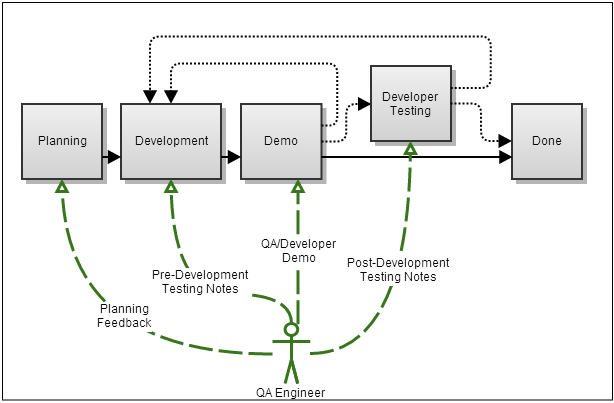}
    \caption{A JIRA software development process and how QA engineers interact with developers prior to releasing a software product.}
    \label{fig:jira}
\end{figure}

Consider an example of an SQA practice inside the Atlassian company, Australia's largest software company with a variety of well-known software products e.g., JIRA Issue Tracking System, BitBucket, and Trello.
Figure~\ref{fig:jira} provides an overview of a JIRA software development process.\footnote{https://www.atlassian.com/blog/inside-atlassian/jira-qa-process}
During this process, a QA engineer has multiple points at which he or she provides feedback into the way the feature is developed and tested---i.e., providing every form of quality improvement guidance for all steps of the software development process from planning to completion.
This process allows for immediate active feedback to ensure that knowledge gained from previous software defects is fed back into the testing notes for future releases to prevent defects in the next iteration.

\begin{figure*}[t]
    \centering
    \includegraphics[width=\textwidth]{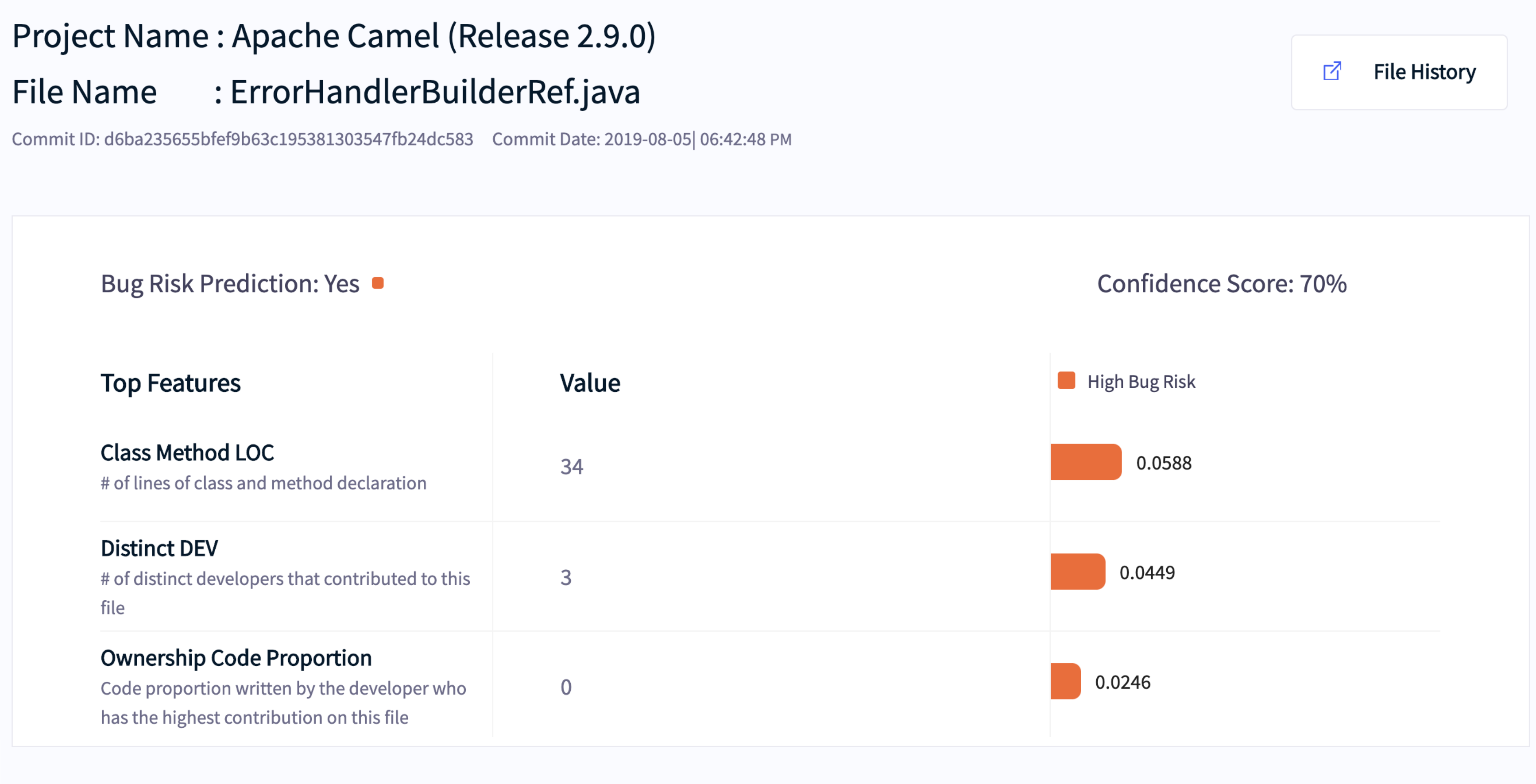}
    \caption{\revisedTextOnly{An example visualization of the Microsoft's Code Defect AI tool (\url{http://codedefectai.azurewebsites.net/}). However, this tool does not suggest what practitioners should do to decrease the risk of having defects, and what practitioners should avoid in order not to increase the risk of having defects. In addition, this tool does not suggest a risk threshold for each metric.}
    }
    \label{Fig:codedefectaiplot}
\end{figure*}

\revisedTextOnly{
\subsection{AI-Driven Defect Prediction and Limitations}


\revised{R2.8}{An AI-driven defect prediction (aka. defect prediction model) is a classification model which is trained on historical data in order to predict if a file is likely to be defective in the future.} 
Defect models serve two main purposes. 
First is \underline{\emph{to predict}}. 
The predictions of defect models can help developers to prioritize their limited SQA resources on the most risky files~\cite{menzies2007data,d2010extensive,tantithamthavorn2018optimization,tantithamthavorn2016automated}.
Therefore, developers can save their limited SQA effort on the most risky files instead of wasting their time on inspecting less risky files.
Second is \underline{\emph{to explain}}.
The insights that are derived from defect models could help managers chart quality improvement plans to avoid the pitfalls that lead to defects in the past~\cite{thongtanunam2016revisiting,Bird2011a,Mcintosh2014}.
For example, if the insights suggest that code complexity shares the strongest relationship with defect-proneness, managers must initiate quality improvement plans to control and monitor the code complexity of that system.

Recently, top software companies have released several commercial AI-driven defect prediction tools.
For example, Microsoft's Code Defect AI,\footnote{\url{https://www.microsoft.com/en-us/ai/ai-lab-code-defect}} Amazon's CodeGuru.\footnote{\url{https://aws.amazon.com/codeguru/}}
Such tools heavily rely on the concept of defect prediction models that have been well-studied in the past decades~\cite{hall2012systematic}.
In particular, Microsoft's Code Defect AI is built on top of the concept of explainable Just-In-Time defect prediction~\cite{jiarpakdee2020xai4se, tantithamthavorn2020explainable}---i.e., explaining the predictions of defect models using a LIME model-agnostic technique~\cite{ribeiro2016should}.
LIME is a model-agnostic technique for explaining the predictions of any AI/ML algorithms.
The crux of Microsoft's Code Defect AI tool is similar to the recent parallel work by Jiarpakdee~\ea~\cite{jiarpakdee2020xai4se}---i.e., extracting several software metrics (e.g., Churn), building a classification model (e.g., random forests), generating a prediction for each file in a commit, and generating an explanation of each prediction using the LIME model-agnostic technique~\cite{ribeiro2016should}.

Figure~\ref{Fig:codedefectaiplot} presents an example visualization of Microsoft's Code Defect AI product for the file \texttt{ErrorHandlerBuilderRef.java} of the Apache Camel Release 2.9.0.
This figure shows that this file is predicted as defective with a confidence score of 70\%.
There are three most important factors that are associated with this prediction as defective, i.e., the number of lines of class and method declaration, the number of distinct developers, and the degree of code ownership.
Thus, these insights can help managers chart quality improvement plans to control for these metrics.
However, there exist the following limitations.

\begin{itemize}
    \item \textbf{First, practitioners still do not know what they should do to decrease the risk of having defects, and what they should avoid to not increase the risk of having defects.} We find that LIME can only indicates what factors are the most important to support the predictions towards defective (G1) and clean (G2) classes, without providing actionable guidance on what should they avoid (G3) and should do (G4) to decrease the risk of having defects.
    \item \textbf{Second, practitioners still do not know a risk threshold for each metric} (e.g., how large is a file size that would be risky? and how small is a file size that would be non-risky?).
\end{itemize}


A lack of these types of guidance and its risk threshold could lead to inefficient and ineffective SQA planning processes.
Such ineffective SQA planning processes could result in the recurrence of software defects, slow project progress, high costs of development, unsatisfactory software products, and unhappy end-users.
To the best of our knowledge, the aforementioned challenges are very significant to the practical applications of defect prediction models, but still remain largely unexplored.

\subsection{A Motivating Scenario for our SQAPlanner}
\label{sec:motivation}

To address the aforementioned challenges, we propose an AI-driven SQAPlanner---i.e., an approach for generating four types of guidance and its risk threshold in the form of rule-based explanation for the predictions of defect models.
Below, we discuss a motivating scenario of how our AI-Driven SQAPlanner could be used in a software development process to assist SQA planning.
}

\smallsection{Without our SQAPlanner} 
Consider Bob who is a QA manager joining a new software development project.
His main responsibility is to apply SQA activities (e.g., code review and testing) to find defects and develop quality improvement plans to prevent them in the next iteration.
However, he has little knowledge of the software projects.
Therefore, he decides to deploy a defect prediction model to guide his QA team about where is the risky areas of source code so his team can effectively allocate the limited effort on this risky area. 
However, Bob still encounters various SQA planning problems during the planning steps to prevent software defects in the next iteration.
In particular, without AI-driven SQA planning tools, he can't understand what are the risky practices and what are the non-risky practices for this team and this project, what are key actions to avoid that increase the risk of having defects, and what are the key actions to do to decrease the risk of having defects.
\emph{A lack of AI-driven SQA planning tools could lead to a failure to develop the most effective SQA plans.}
Ultimately, this results in the recurrence of software defects, slow project progress, and high costs of software development, unsatisfactory software products, and unhappy end-users.

\smallsection{With our SQAPlanner} 
Now consider that Bob adopts our AI-driven SQAPlanner tool. 
In particular, given a file that is predicted as defective by defect prediction models, our SQAPlanner can further generate rule-based explanations to better understand what are key risky practices, non-risky practices, actions to avoid that increase the risk of defects, and actions to do to decrease the risk of having defects for that file.
Bob can use our SQAPlanner to make \emph{data-informed decisions} when developing SQA plans.
This could result in more optimal SQA plans, leading to higher quality of software systems, less number of software defects, lower costs of software development, satisfactory software products, and happy end-users.


\subsection{The Design Rationale for the Four Types of Guidance}
\label{sec:guidance}

First, we propose to generate the guidance in the form of rule-based explanations, since our recent work~\cite{jiarpakdee2020xai4se} found that decision trees/rules are the most preferred representation of explanations by software practitioners as they involve logic reasoning that they are familiar with.
Formally, a rule-based explanation ($e$) is an association rule $e = \{r = p \Rightarrow q \}$ that describes the association between $p$ (a Boolean condition of feature values (i.e., antecedent, left-hand-side, LHS)) and $q$ (the consequence (i.e., consequent, right-hand-side, RHS)) for the decision value $y = f'(x)$.
\revised{R2.5, R3.2}{In this paper, we use an arrow ($\xLongrightarrow{\text{associate}}$) to describe the association between the Boolean condition ($p$) of feature values for a file and the predictions ($q$) towards a \{DEFECT,CLEAN\} class. 
Note that an association in general doesn't mean that there is a causal relationship.}

\revised{R1.5, R2.2}{
Second, motivated by the limitations of Microsoft's Code Defect AI tool (see Figure~\ref{Fig:codedefectaiplot}), we hypothesize that the following four types of guidance (G) that are presented in a form of rule-based explanations are beneficial to guide practitioners when developing SQA plans.
Below, we present the definition, the motivation, and an example of the four types of guidance.

\begin{enumerate}[{\bf G1:}]
\item \textbf{Risky current practices that lead the defect model to predict a file as defective} are needed to help practitioners understand what current practices are problematic. 
For example, an association rule of $\{\mathrm{LOC}>100\}\xLongrightarrow{\text{associate}}\mathrm{DEFECT}$ indicates that a file with LOC greater than 100 is associated with the predictions towards a defective class.
Thus, practitioners should consider decreasing the LOC to less than 100, as this may likely decrease the risk of having defects.

\item \revised{R2.3}{\textbf{Non-risky current practices that lead the defect model to predict a file as clean} are needed to help practitioners understand what current practices contribute towards a low risk of having defects. 
For example, an association rule of $\{\mathrm{Ownership}>0.8\} \xLongrightarrow{\text{associate}} \mathrm{CLEAN}$ indicates that a file with an ownership value greater than 0.8 is associated with the predictions towards a clean class.
Thus, practitioners should consider maintaining or increasing the ownership value to more than 0.8 to potentially decrease the risk of having defects.}

\item \revised{R1.5, R2.2, R2.4}{\textbf{Potential practices to avoid to not increase the risk of having defects}} are needed to help practitioners understand which currently not implemented practices to avoid to not increase the risk of having defects. 
For example, an association rule of $ \{ \mathrm{MinorDeveloper}>0\} \xLongrightarrow{\text{associate}} \mathrm{DEFECT}$ indicates that a file with a number of minor developers of greater than 0 is associated with the predictions towards a defective class.
Thus, practitioners should avoid increasing the number of minor developers to greater than zero to not increase the risk of having defects.

\item \textbf{Potential practices to follow to decrease the risk of having defects} are needed to help practitioners understand which practices to newly implement to decrease the risk of having defects. 
For example, an association rule of $\{ \mathrm{RatioCommentToCode} > 0.6\} \xLongrightarrow{\text{associate}} \mathrm{CLEAN}$ indicates that a file with a proportion of comments to code that is larger than 60\% is associated with the predictions towards the clean class.
Thus, practitioners should consider increasing the proportion of comments to code to greater than 60\% to decrease the risk of having defects.

\end{enumerate}
}

\begin{figure}[t]
	\centering
	\includegraphics[width=\columnwidth]{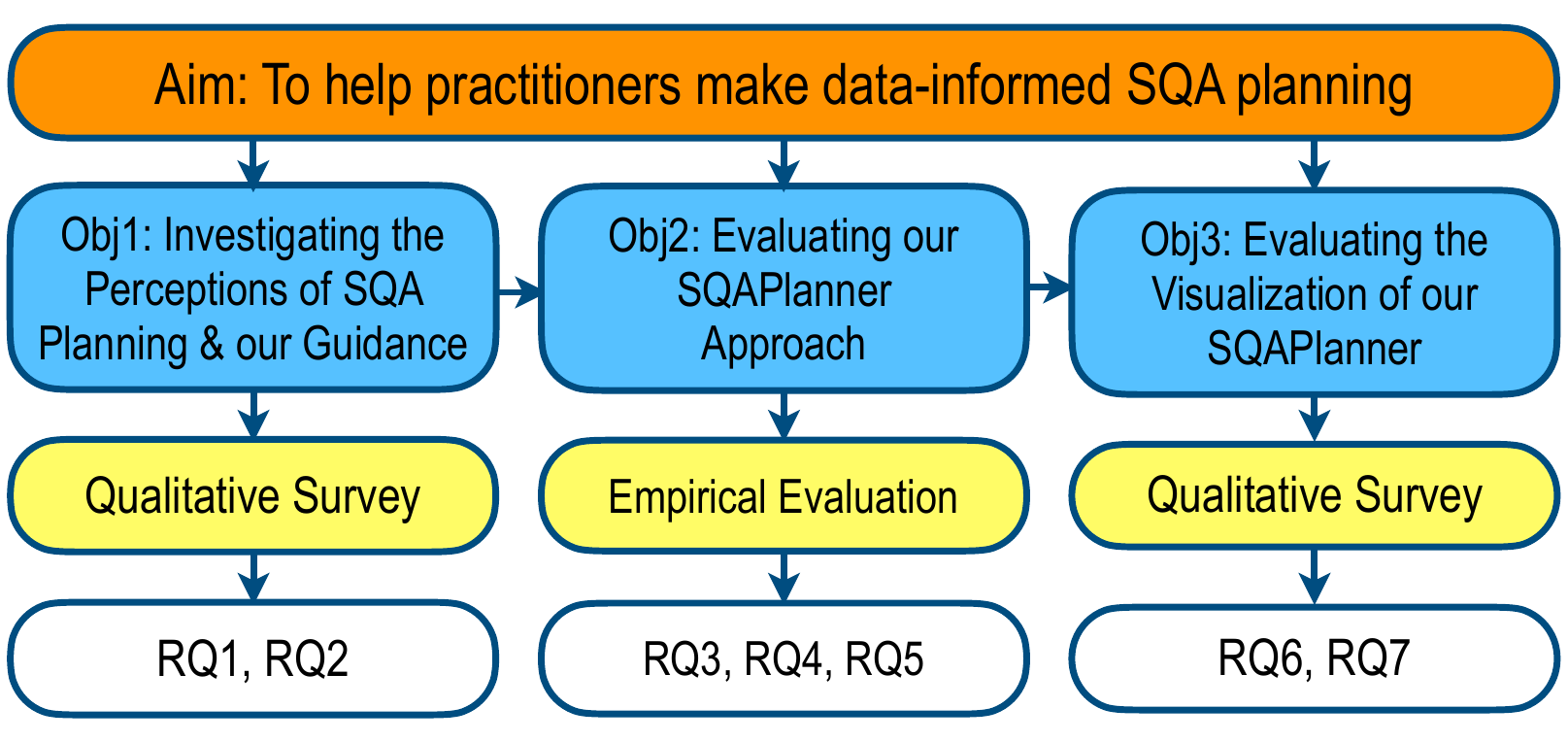}
	\caption{\revisedTextOnly{An overview of our study design and research questions.}}
	\label{fig:studyoverview}
\end{figure}

\revisedTextOnly{

\section{Study Design and Research Questions}
\label{sec:rqs}

\revisedTextOnly{
In this paper, we aim to help practitioners make data-informed SQA planning by providing guidance on (1) what practitioners should do to decrease the risk of having defects and (2) what practitioners should avoid in order not to increase the risk of having defects with (3) a risk threshold in the form of rule-based explanations for the predictions of defect prediction models.
To achieve this aim, we design our case study according to the following objectives (see Figure~\ref{fig:studyoverview}):}

\textbf{Objective 1---Investigating the practitioners' perceptions of SQA planning and the proposed four types of guidance.}
SQA planning activities are important in software development processes (e.g., to define initial software development policies), but often vary from organization to organization~\cite{farooqui2017survey}. 
However, there exist no empirical studies that investigate how practitioners perceive the importance of SQA planning activities in their organization and what are their key challenges.
Thus, we formulate the following research question:

\begin{itemize}
    \item \textbf{(RQ1) \rqi}
\end{itemize}

One of the most important SQA planning activities is to define development policies and their associated risk thresholds~\cite{galin2018software}.
Such development policies will be later enforced for the whole team to ensure the highest quality of software systems (e.g., the maximum file size, the maximum code complexity, the minimum code to comment ratio, and the minimum degree of code ownership).
Such policies are essential to improve software quality and software maintainability.
Recently, Microsoft's Code Defect AI tool has been released to the public where the crux of this tool is defect prediction models.
However, Figure~\ref{Fig:codedefectaiplot} shows that such tool only indicates the importance scores of features that are generated by LIME, which are still far from actionable.
That means LIME only indicates what factors are the most important to support the predictions towards defective (G1) and clean (G2) classes, but do not actually guide developers what should they avoid (G3) and should do (G4) to decrease the risk of having defects.
We hypothesize that our proposed four types of guidance that are presented in a form of rule-based explanation would be more actionable to guide practitioners when developing SQA plans.
Thus, we formulate the following research question:

\begin{itemize}
    \item \textbf{(RQ2) \rqii}
\end{itemize}

\textbf{Objective 2---Developing and Evaluating our AI-Driven SQAPlanner Approach.}
To address the practitioners' challenges of SQA planning and the limitations of Microsoft's Code Defect AI tool, we propose SQAPlanner to help practitioners make data-informed decisions when developing SQA plans.
First, SQAPlanner develops a defect prediction model to generate a prediction.
Then, SQAPlanner generates a rule-based explanation of the prediction to provide actionable guidance.
However, there are different local rule-based model-agnostic techniques for generating explanations in the eXplainable AI (XAI) domain available (e.g., Anchor~\cite{ribeiro2018anchors} and LORE~\cite{guidotti2018local}).
Thus, it remains unclear whether our SQAPlanner outperforms the state-of-the-art rule-based model-agnostic techniques.
Therefore, we conduct an empirical study to evaluate our approach and compare with the baseline techniques.
Thus, we formulate the following research questions.

\begin{itemize}
    \item \textbf{(RQ3) \rqiii}
    \item \textbf{(RQ4) \rqiv}
    \item \textbf{(RQ5) \rqv}
\end{itemize}

\textbf{Objective 3---Developing the Visualization of SQAPlanner and Investigating the Practitioners' Perceptions.}
While the rule-based explanations of our SQAPlanner are designed to help practitioners understand the logic behind the predictions of defect models, such rule-based explanations may not be immediately actionable and easily understandable by practitioners.
Thus, we develop a proof-of-concept by translating the rule-based explanations of the actionable guidance into human-understandable explanations.
The visualization of our SQAPlanner is designed to provide the following key information: (1) the list of guidance that practitioners should follow and should avoid; (2) the actual metric values of that file; and (3) the risk threshold and range values for practitioners to follow to mitigate the risk of having defects.
Then, we conduct a post-validation qualitative survey with practitioners to evaluate their perceptions of the visualization of our {SQAPlanner} when comparing to the existing visualization of Microsoft's Code Defect AI (see Figure~\ref{Fig:codedefectaiplot}).
Thus, we formulate the following research questions:

\begin{itemize}
    \item \textbf{(RQ6) \rqvi}
    \item \textbf{(RQ7) \rqvii}
\end{itemize}

}

\begin{table*}[htb!]
\renewcommand{\arraystretch}{1.2}
\centering
\caption{(RQ1 and RQ2) A summary of the agreement percentage, the disagreement percentage, and the agreement factor for the practitioners' perception of SQA planning activities and our proposed four types of guidance.}
\label{survey-table}

\begin{footnotesize}
\begin{tabular}{c||l|c|c|c}
\hline
\textbf{Dimension}                       & \textbf{Statement} & \textbf{\%Agreement} & \textbf{\%Disagreement} & \textbf{Agreement Factor}      \\ \hline\hline
\multirow{1}{*}{{(RQ1) Perceived importance}}     & \multirow{4}{*}{{SQA planning activities}}     &	86\%	&	6\%	&	14.33          \\ \cline{1-1} \cline{3-5}
\multirow{1}{*}{{(RQ1) Being used in practice}}     &     &	70\%	&	10\%	&	7.00          \\ \cline{1-1} \cline{3-5}
\multirow{1}{*}{{(RQ1) Perceived time-consuming}}     &     &	66\%	&	10\%	&	6.60        \\ \cline{1-1} \cline{3-5}
\multirow{1}{*}{{(RQ1) Perceived difficulty}}     &     &	58\%	&	24\%	&	2.42        \\ \hline \hline
\multirow{8}{*}{(RQ2) Perceived usefulness}     & \begin{tabular}[c|]{@{}l@{}}G1: Risky current practices that lead\\the defect model to predict a file as defective\end{tabular}   &	82\%	&	6\%	&	13.67          \\ \cline{2-5}
& \begin{tabular}[c|]{@{}l@{}}G2: Non-risky current practices that lead\\the defect model to predict a file as clean\end{tabular}                       &	64\%	&	10\%	&	6.40                                                      \\ \cline{2-5}
& \begin{tabular}[c|]{@{}l@{}}G3: Potential practices to avoid to\\not increase the risk of having defects\end{tabular}                 &	52\%	&	20\%	&	2.60                                                      \\ \cline{2-5}
& \begin{tabular}[c|]{@{}l@{}}G4: Potential practices to follow to decrease\\the risk of having defects
\end{tabular}                  &	80\%	&	8\%	&	10.00                                                     \\ \hline
\multirow{8}{*}{(RQ2) Perceived importance}     & \begin{tabular}[c|]{@{}l@{}}G1: Risky current practices that lead\\the defect model to predict a file as defective\end{tabular}   &	64\%	&	10\%	&	6.40          \\ \cline{2-5}
& \begin{tabular}[c|]{@{}l@{}}G2: Non-risky current practices that lead\\the defect model to predict a file as clean\end{tabular}                       &	60\%	&	10\%	&	6.00                                                    \\ \cline{2-5}
& \begin{tabular}[c|]{@{}l@{}}G3: Potential practices to avoid to\\not increase the risk of having defects\end{tabular}                  &	64\%	&	24\%	&	2.67                                                     \\ \cline{2-5}
& \begin{tabular}[c|]{@{}l@{}}G4: Potential practices to follow to decrease\\the risk of having defects
\end{tabular}                 &	82\%	&	6\%	&	13.67                                                    \\ \hline
\multirow{8}{*}{(RQ2) Willingness to adopt}     & \begin{tabular}[c|]{@{}l@{}}G1: Risky current practices that lead\\the defect model to predict a file as defective\end{tabular}   &	74\%	&	12\%	&	6.17          \\ \cline{2-5}
& \begin{tabular}[c|]{@{}l@{}}G2: Non-risky current practices that lead\\the defect model to predict a file as clean\end{tabular}                       &	66\%	&	12\%	&	5.50                                                    \\ \cline{2-5}
& \begin{tabular}[c|]{@{}l@{}}G3: Potential practices to avoid to\\not increase the risk of having defects\end{tabular}                  &	52\%	&	22\%	&	2.36                                                     \\ \cline{2-5}
& \begin{tabular}[c|]{@{}l@{}}G4: Potential practices to follow to decrease\\the risk of having defects\end{tabular}                &	72\%	&	12\%	&	6.00                                                      \\ \hline\hline
\end{tabular}
\end{footnotesize}
\end{table*}

\section{Practitioners' Perceptions on SQA Planning and the Four Types of Guidance}\label{sec:survey}

In this section, we aim to investigate the practitioners' perceptions of (1) the SQA planning activities (RQ1) and (2) the proposed four types of guidance to support SQA planning (RQ2).
Below, we describe the approach and present the results.

\subsection{Approach} 

To investigate practitioners' perceptions of SQA planning activities and their feedback on our proposed four types of data-driven guidance to support such activities, we conducted a survey study with 50 software practitioners. 
As suggested by Kitchenham and Pfleeger~\cite{kitchenham2008personal}, we considered the following steps when conducting our study: (1) design and develop a survey, (2) evaluate a survey, (3) recruit and select participants, (4) verify data, and (5) analyse data.
We describe each step below.

\textbf{(Step 1) Design and develop a survey.}
We first devised the concept of data-driven software quality assurance (SQA) planning with respect to the 4 types of rules generated by our approach.
We then wanted to investigate practitioners' perceptions along 4 dimensions, i.e., perceived importance, being used in practice, would it be time-consuming, and what are key difficulties.
We designed our survey as a cross-sectional study where participants provide their responses at one fixed point in time. 
The survey consists of 16 closed-ended questions and 4 open-ended questions. 
For closed-ended questions, we use agreement and evaluation ordinal scales. 
To mitigate any inconsistency of the interpretation of numeric ordinal scales, we labeled each level of the ordinal scales with words as suggested by Krosnick~\cite{krosnick1999survey} (e.g., strongly disagree, disagree, neutral, agree, and strongly agree).
The format of the survey is an online questionnaire where we use an online questionnaire service as provided by Google Forms.
When accessing the survey, each participant is provided with an explanatory statement that describes the purpose of the study, why the participant is chosen for this study, possible benefits and risks, and confidentiality.
The survey takes approximately 15 minutes to complete and is anonymous.

\textbf{(Step 2) Evaluate a survey.} 
We carefully evaluated the survey via pre-testing~\cite{litwin1995measure} to assess the reliability and validity of the survey. 
We revised the evaluation process to identify and fix potential problems (e.g., missing, unnecessary, or ambiguous questions) until reaching a consensus.
Finally, the survey has been rigorously reviewed and approved by the Monash University Human Research Ethics Committee (MUHREC ID: 22542).

\textbf{(Step 3) Recruit and select participants.}
The target population of the survey is software practitioners. 
To reach the target population, we used a recruiting service provided by the Amazon Mechanical Turk to recruit 50 participants as a representative subset of the target population. 
We use the participant filter options of "Employment Industry - Software \& IT Services" and "Job Function - Information Technology" to ensure that the recruited participants are valid samples representing the target population. 
We pay 6.4 USD as a monetary incentive for each participant~\cite{smith2013improving,edwards2002increasing}.

\revised{R2.7}{\textbf{(Step 4) Verify data.}
To verify our survey response data, we manually read all of the open-question responses to check the completeness of the responses i.e., whether all questions were appropriately answered.
We excluded 11 responses that are missing and are not related to the questions.
In the end, we had a set of 989 responses.
We summarized and presented the results of closed-ended responses in a Likert scale with stacked bar plots, while we discussed and provided examples of open-ended responses.}

\begin{figure}[t]
	\centering
	\includegraphics[width=\columnwidth, scale=2,trim={20 0 0 0}]{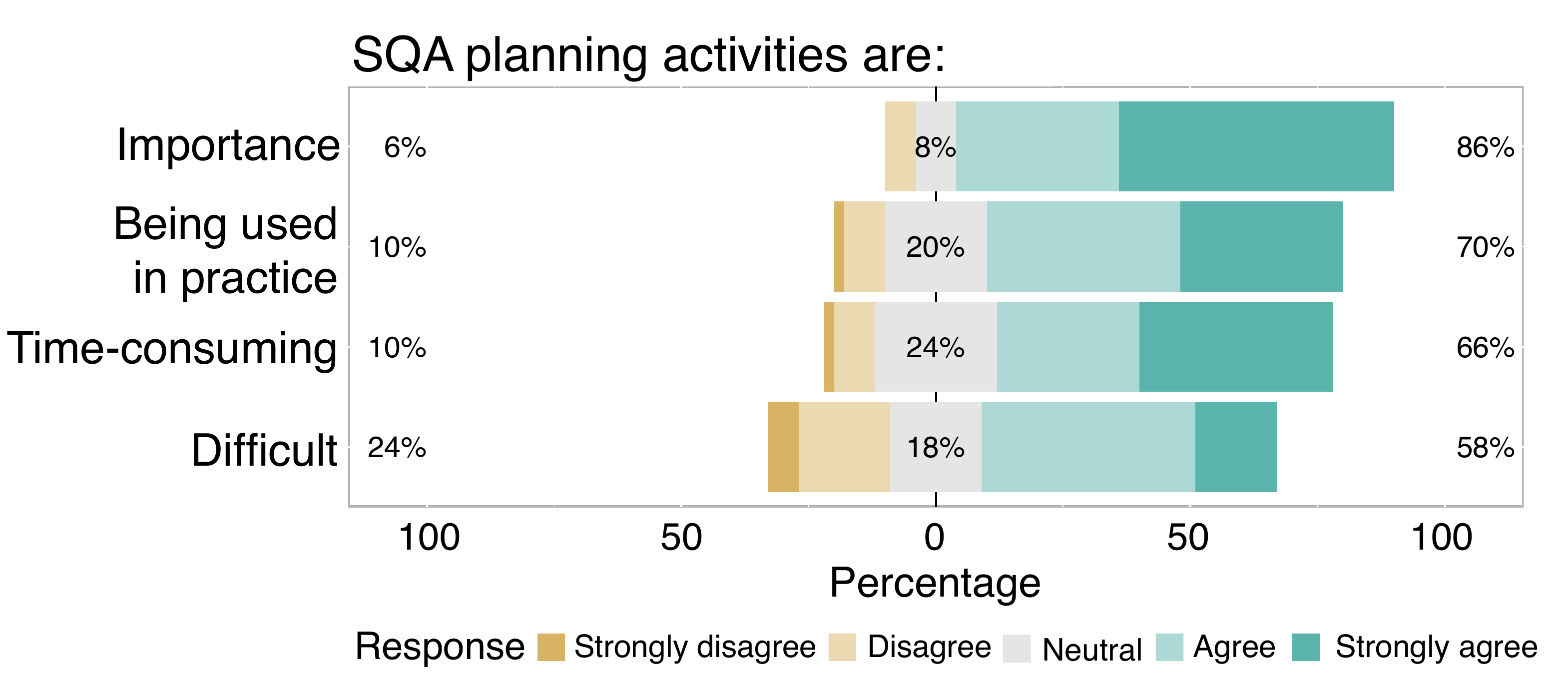}
	\caption{(RQ1) The likert scores of the practitioners' perceptions of SQA planning along four dimensions i.e., importance, being used in practice, time-consuming, and difficulty.}
	\label{fig:rq4-results}
\end{figure}

\begin{figure}[t]
\centering
\begin{subfigure}[t]{\hsize}
\includegraphics[width=\linewidth]{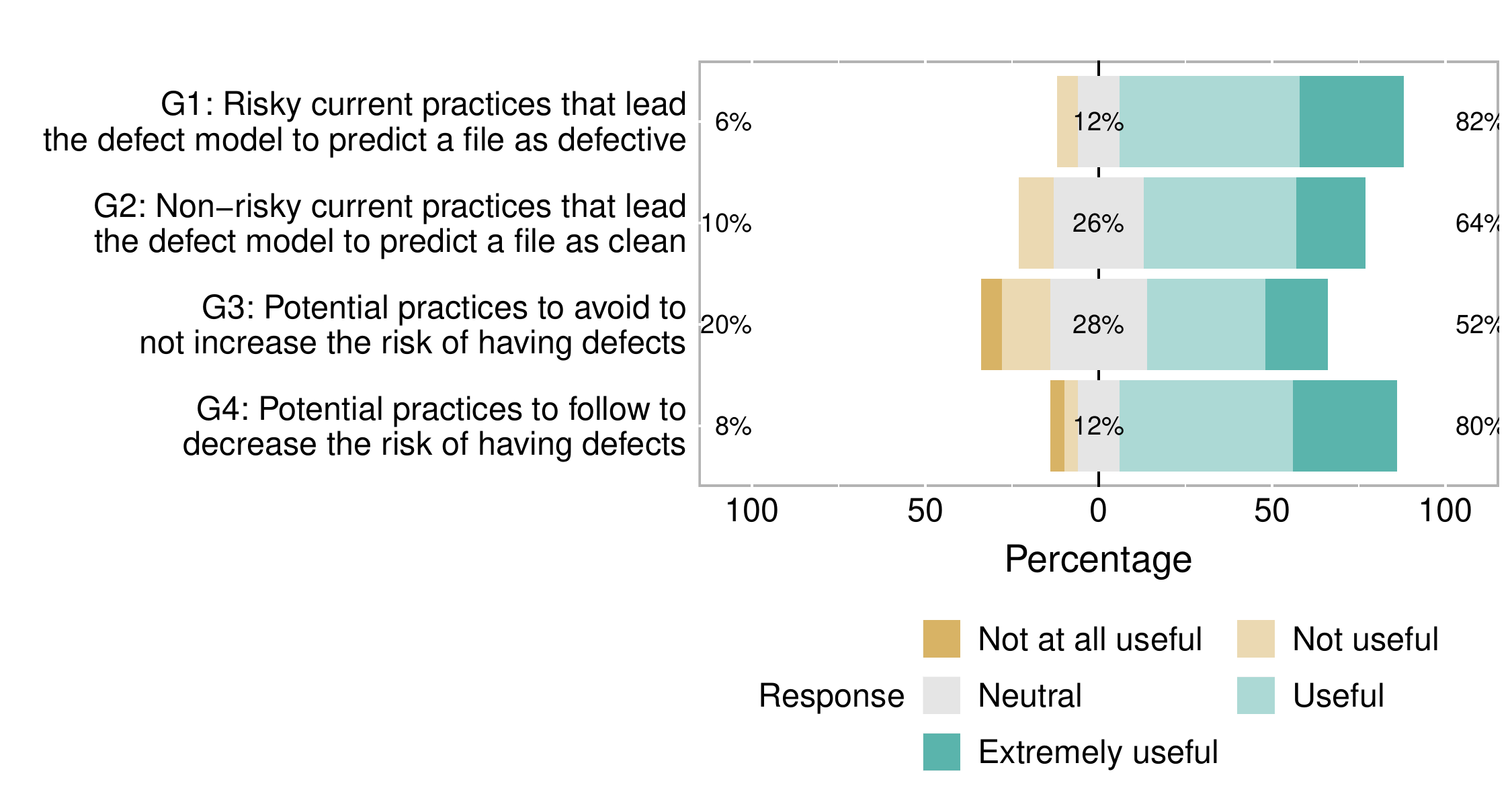}
\caption{Perceived usefulness}
\label{fig:rq5-results-1}
\end{subfigure} 
\hfill
\begin{subfigure}[!t]{\hsize}
\includegraphics[width=\linewidth]{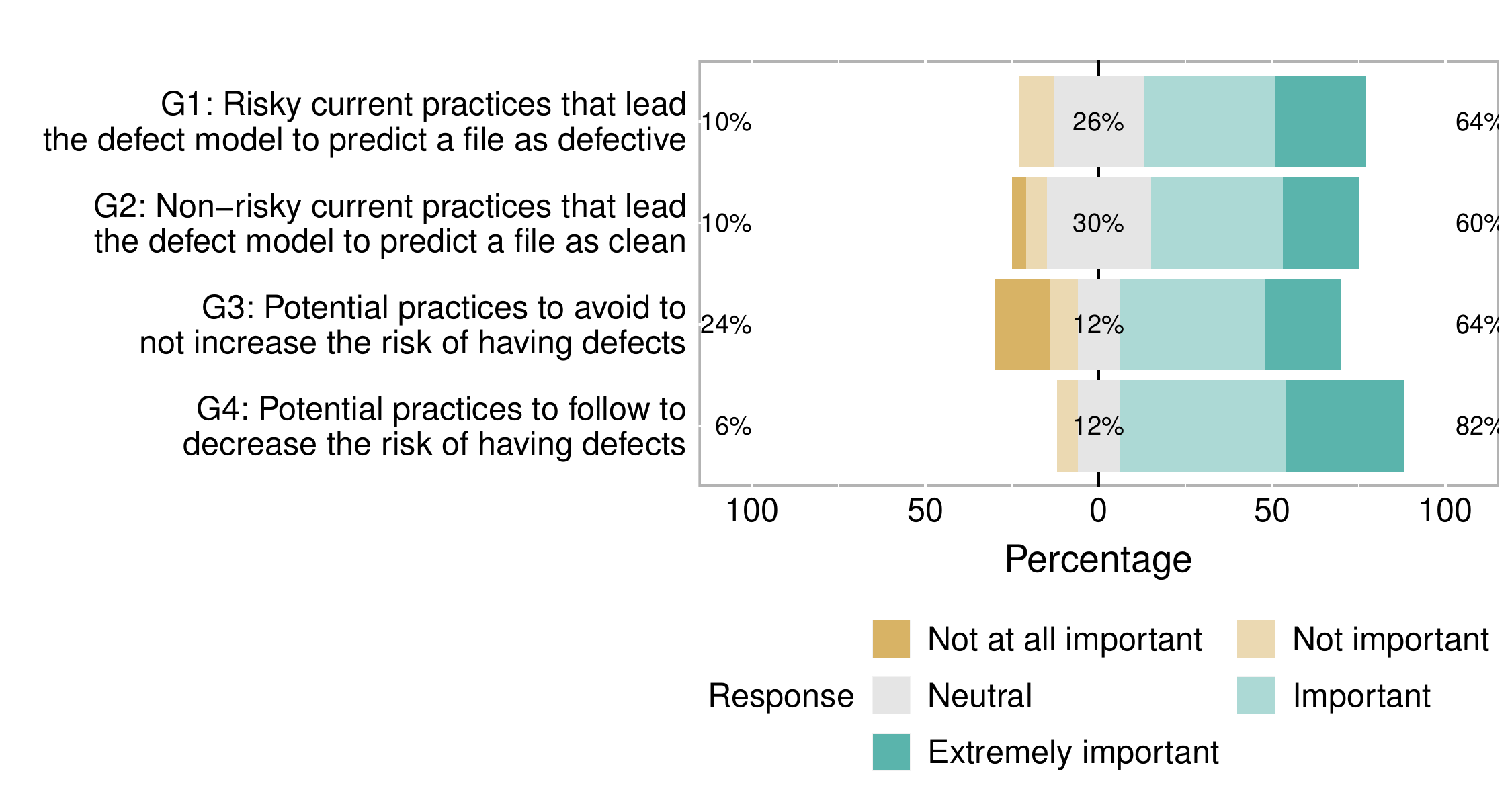}
\caption{Perceived importance}
\label{fig:rq5-results-2}
\end{subfigure} 
\hfill
\begin{subfigure}[!t]{\hsize}
\includegraphics[width=\linewidth]{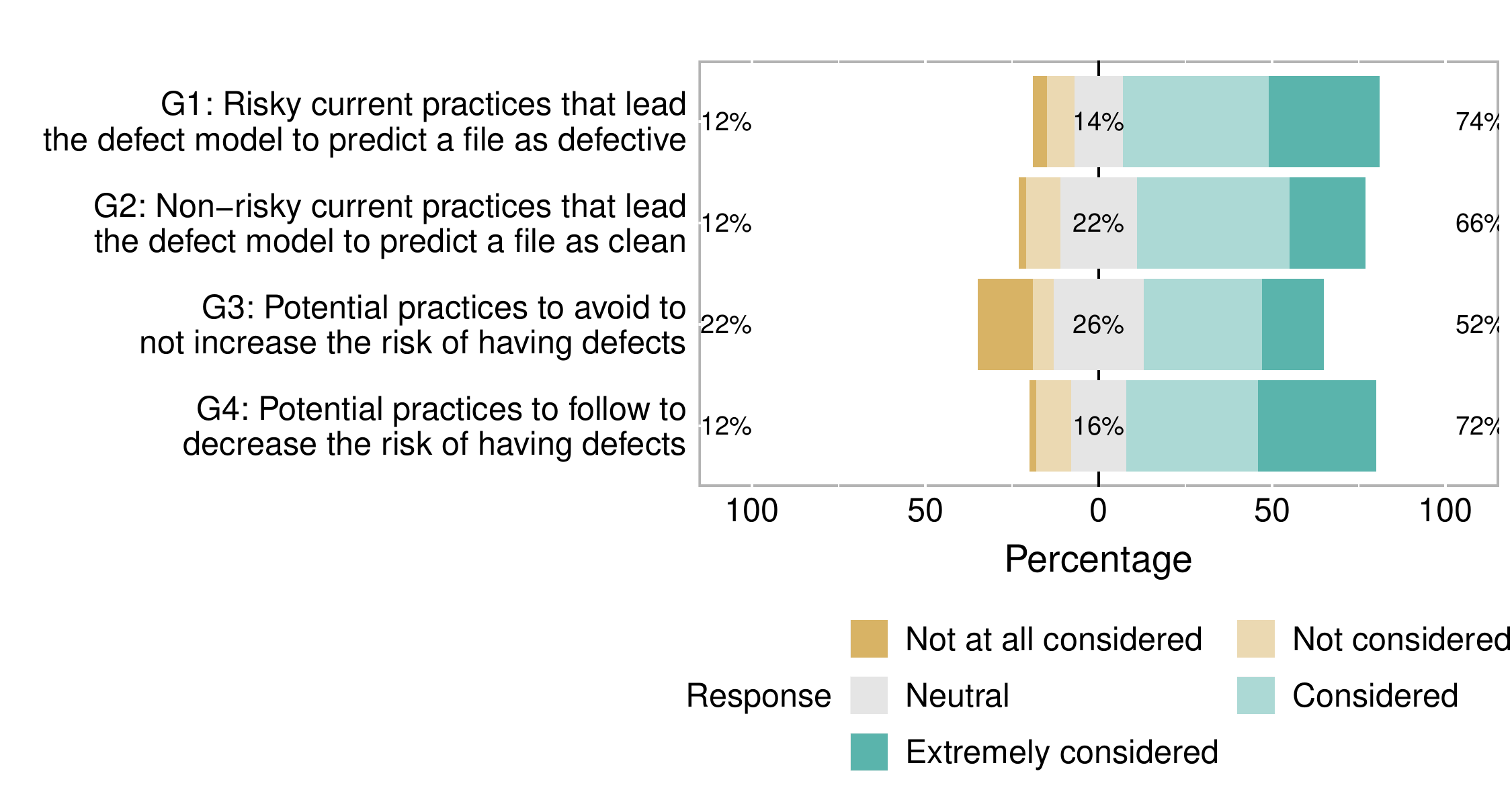}
\caption{Willingness to adopt}
\label{fig:rq5-results-3}
\end{subfigure} 
\caption{(RQ2) The likert scores of the perceived usefulness, the perceive importance, and the willingness to adopt of the respondents for each proposed guidance.}
\label{fig:rq5-results}
\end{figure}  

\textbf{(Step 5) Analyse data.}
We manually analysed the responses of the open-ended questions to extract in-depth insights. 
For closed-ended questions, we summarise and present key statistical results.
We compute the agreement and disagreement percentage of each closed-ended question.
The agreement percentage of a statement is the percentage of respondents who strongly agree or agree with a statement (\emph{\% strongly agree + \% agree}), while the disagreement percentage of a statement is the percentage of respondents who strongly disagree or disagree with a statement (\emph{\% strongly disagree + \% disagree}).
We also compute an agreement factor of each statement as suggested by Wan~\ea~\cite{wan2018perceptions}.
The agreement factor is a measure of agreement between respondents, which is calculated for each statement using the following equation: (\emph{\% strongly agree + \% agree})/(\emph{\% strongly disagree + \% disagree}).
High values of agreement factors indicate a high agreement of respondents to a statement.
The agreement factor of 1 indicates that the numbers of respondents who agree and disagree with a statement are equal.
Finally, low values of agreement factors indicate that a high disagreement of respondents to a statement.

\subsection{Respondent Demographics}

The demographics of our 50 practitioner survey respondents are as follows:
 
\begin{itemize}
    \item Country of Residence: India (58\%) and US (36\%)
    \item Roles: developers (50\%), managers (42\%), and others (8\%)
    \item Years of Professional Experience: less than 5 years (26\%), 6--10 years (38\%), 11--15 years (22\%), 16--20 years (12\%), and more than 25 years (2\%)
    \item Programming Language: Java (44\%), Python (30\%), C/C++/C\# (28\%), and JavaScript (12\%)
    \item Use of Static Analysis Tools: Yes (62\%) and No (38\%)
\end{itemize}

These demographics indicate that the responses are collected from practitioners who reside in various countries, have a range of roles, varied years of experience, and varied programming language expertise. This indicates that our findings are likely not bound to specific characteristics of practitioners.

\subsection{Results}

\subsection*{(RQ1) \rqi}


\smallsection{Results}
\textbf{For SQA planning activities, 86\% of the respondents perceive as important and 70\% perceived as being used in practice. However, 66\% perceived as time-consuming and 58\% perceived as difficult.}
Figure~\ref{fig:rq4-results} shows the distributions of likert scores of the practitioners' perceptions of SQA planning activities.
The survey results show that SQA planning activities are perceived as important by 86\% of the respondents, and are being used in practice by 70\% of the respondents. 
However, they are perceived as time-consuming by 66\% of the respondents and as difficult to do by 58\% of the respondents.
Table~\ref{survey-table} also shows that the agreement factor of all studied dimensions of SQA planning activities are of above 1 with the values of 2.42 - 14.33.
This indicates that most respondents agree (while having very few respondents who disagree) that SQA planning activities are important, being used in practice, time-consuming, and difficult. 

Respondents described that some of the SQA planning activities in their organisations involve human heuristics in decision-making.
For example, they used documentation and review checklists~\cite{chong2021assessing} (e.g., R34: \emph{``Lessons learnt from projects are documented and common mistakes are included in review checklists to ensure that they are not repeated.''}), and team meetings (e.g., R10: \emph{``team meetings, brainstorm, and in house system''}, and R48: \emph{``... through step by step manual processes working together in a core team''}).
These findings indicates that a data-informed SQA planning tool is needed to support QA teams make better data-informed decision- and policy-making.


\subsection*{(RQ2) \rqii}

\smallsection{Results}
\textbf{Both (G1) the guidance on risky practices that lead a model to predict a file as defective and (G4) the guidance on the practices to follow to decrease the risk of having defects are perceived as among the most useful, most important, and most considered willingness to adopt by the respondents.}
Figure~\ref{fig:rq4-results} shows the likert scores of the practitioners’ perceptions of SQA planning along four dimensions i.e., importance, being used in  practice, time-consuming, and difficulty.
The survey results show that all types of guidance are perceived as useful by 52\%-80\% of the respondents, important by 60\%-82\% of the respondents, and considered willing to adopt by 52\%-72\% of the respondents.
Similar to RQ1, we observed that the values of agreement factor for all of the proposed guidance are higher than 1 for all of the studied dimensions.
This suggests that most respondents agree (while having a very few of those who disagree) that all proposed guidance are useful, important, and willing to adopt these four types of guidance.

Respondents provided positive feedback of our proposed four types of guidance since these types of guidance can help with SQA planning (e.g., R37: \emph{``It allows the QA team who might not necessarily know the changes that have gone into each program to focus their energy on the most risky components, programs, or functionalities. It also gives managers a great view of the risks involved and how it could potentially be reduced or mitigated.''}).

\revisedTextOnly{However, some respondents raise critical concerns related to the potential negative impact on the development process made by these four types of guidance.}
\revised{R1.6}{For example, cost of implementation and internal resistance (e.g., R27: \emph{``Some extra time spent improving the process. Needing to implement the process including training. Employee resistance to adoption.''}), and lax development practice (e.g., R30: \emph{``Sometimes we get too reliant on the automated processes and other things slip through ...''}).}


%


\section{Our AI-Driven SQAPlanner Approach}
\label{sec:methodology}

\begin{figure}[t]
	\centering
	\includegraphics[width=\columnwidth]{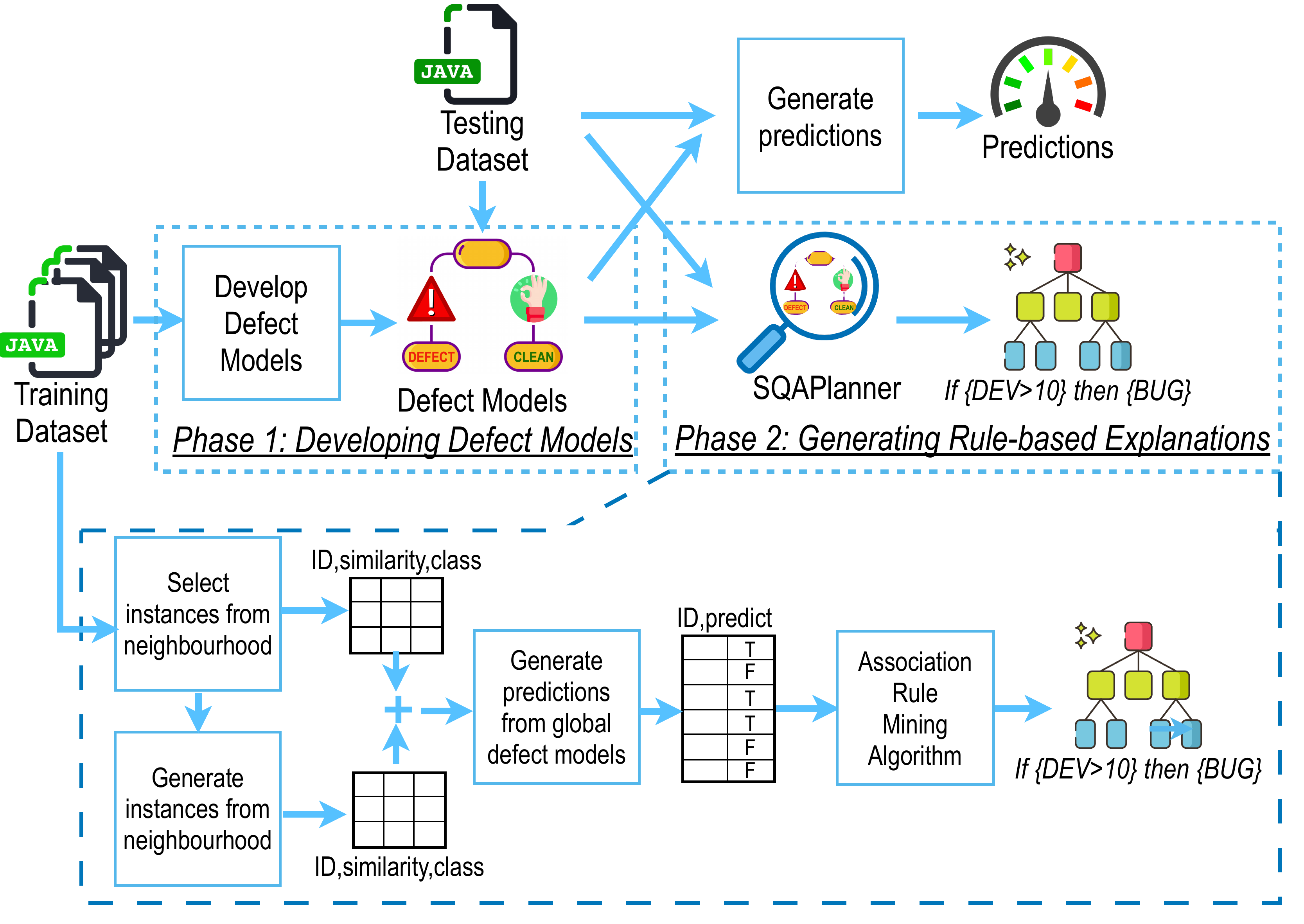}
	\caption{An overview diagram of our SQAPlanner to generate four types of guidance in the form of rule-based explanations for each file.}
	\label{fig:framework}
\end{figure}

Our SQAPlanner consisted of two major phases: (1) developing defect prediction models; and (2) generating four types of guidance using a local rule-based model-agnostic technique to explain the predictions of defect models.
Figure~\ref{fig:framework} presents an overview workflow of our SQAPlanner approach.

\subsection{Phase 1: Developing Defect Prediction Models}


There is a plethora of classification techniques that have been used to develop defect prediction models~\cite{ghotra2015revisiting,tantithamthavorn2016automated,hall2012systematic}.
We first select the following five classification techniques, i.e., Decision Trees (DT), Logistic Regression (LR), multi-layer Neural Network (NN), Random Forest (RF), and Support Vector Machine (SVM).
These classification techniques are popularly-used in defect prediction studies.
Since the performance of defect prediction models may vary depending on the studied datasets, we first conduct a preliminary analysis to identify the most accurate classification techniques for our study.
We use the implementation of the selected five classification techniques provided by the scikit-learn Python package.

For each training dataset, we build defect prediction models using all of the 65 software metrics (see Table~\ref{tab:staticmetrics} and Table~\ref{tab:processmetrics}).
\revised{R1.4}{
To ensure that our experiment is strictly-controlled and fair across the studied classification techniques, we use the default setting of the classification techniques provided by the scikit-learn Python package, do not apply feature selection techniques, and do not apply class rebalancing techniques.
This setting will ensure that the results are not bound to (i.e., not sensitive to) the randomization of the non-deterministic optimization algorithms~\cite{tantithamthavorn2018optimization}, feature selection algorithms~\cite{jiarpakdee2018impact}, and class rebalancing algorithms~\cite{tantithamthavorn2018impact}.
}
Then, we evaluate the performance of each classification technique using testing datasets.
Then, we measure the predictive ability of defect models using an Area Under the Receiver Operating Characteristic Curve (AUROC or AUC).
AUC measures the ability to distinguish defective and clean files. 
The values of AUC range from 0 to 1.
The AUC value of 0 is considered the worst performance, the AUC value of 0.5 is considered as merely random guessing, and the AUC value of 1 is considered the best performance~\cite{Hanley1982-uo}.  

\revisedTextOnly{
Then, we use the Non-Parametric Scott-Knott ESD test (Version 3.0) to find the classification techniques that perform best across our studied datasets. 
We chose the Non-Parametric Scott-Knott ESD test, since it does not produce overlapping groups like other post-hoc tests (e.g., Nemenyi’s test)~\cite{tantithamthavorn2017empirical} and it does not require the assumptions of normal distributions, homogeneous distributions, and the minimum sample size. 
The Non-Parametric ScottKnott ESD test is a multiple comparison approach that leverages a hierarchical clustering to partition the set of median values of techniques (e.g., medians of variable importance scores, medians of model performance) into statistically distinct groups with non-negligible difference.
The mechanism of the Non-Parametric Scott-Knott ESD test consists of 2 steps: (Step 1) Find a partition that maximizes the median of each distribution between groups using the non-parametric Kruskal-Wallis test with Chi-square statistics. (Step 2) Split the distributions into two groups or merging into one group using the non-parametric Cliff $|\delta|$ effect size.}
\revised{R1.7}{
The implementation of the Non-Parametric ScottKnott ESD test is available in the ScottKnott ESD R package (Version 3.0).\footnote{http://github.com/klainfo/ScottKnottESD}
}

\begin{figure}[t]
	\centering
	\includegraphics[width=0.8\linewidth]{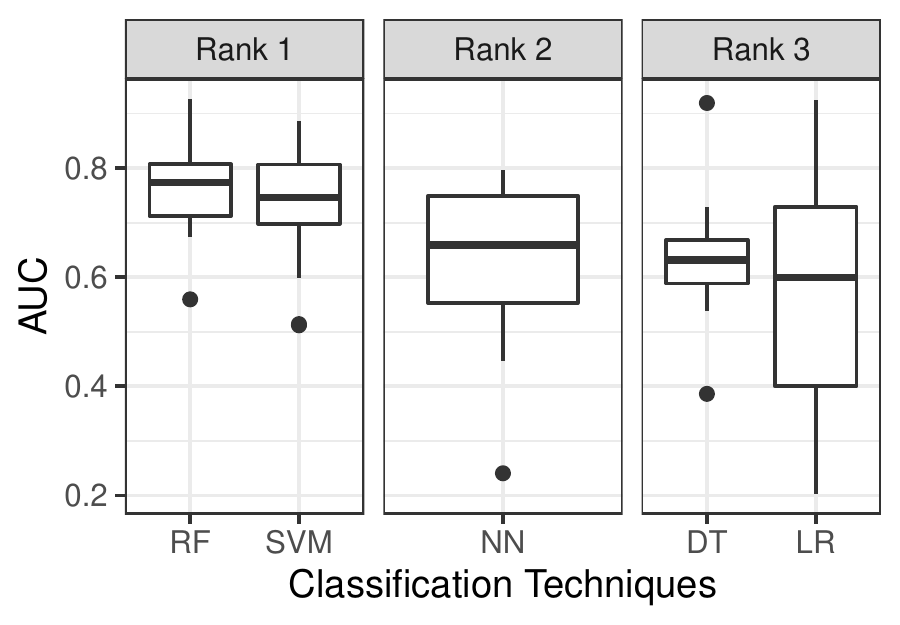}
	\caption{The Non-Parametric Scott-Knott ESD ranking of the studied classification techniques with the distribution of the AUC values.}
	\label{fig:bestmodel_auc}
\end{figure}

\textbf{Random Forest is the most accurate studied classification technique with a median AUC value of 0.77.}
Figure~\ref{fig:bestmodel_auc} presents the Scott-Knott ESD ranking of the studied classification techniques with the distribution of the AUC values.
We find that other classification techniques achieve a median AUC value of 0.74, 0.63, 0.65, and 0.59 for SVM, DT, NN, LR, respectively.
Finally, the ScottKnottESD test confirms that random forests statistically outperforms other classification techniques.
\revisedTextOnly{
For the rest of the paper, we focus on the random forest models due to the following reasons:}
\revised{R2.12}{
\begin{itemize}
    \item Random Forest is one of the most accurate studied classification techniques for our case study and is less sensitive to parameter settings~\cite{tantithamthavorn2016automated,tantithamthavorn2018optimization};
    \item Random Forest is a classification technique that is to a certain degree explainable with its own built-in feature importance techniques (e.g., gini importance and permutation importance)~\cite{breiman2011package,jiarpakdee2018impact,jiarpakdee2018autospearman, jiarpakdee2020xai4se}. Since SVM does not have its own built-in feature importance techniques, we excluded SVM from our analysis; and
    \item Random Forest is a classification technique that is robust to overfitting~\cite{tantithamthavorn2016automated}, outliers~\cite{tantithamthavorn2018impact}, and class mislabelling~\cite{tantithamthavorn2015icse}.
\end{itemize}
}

\subsection{Phase 2: Generating Four Types of Guidance Using a Local Rule-based Model-agnostic Technique}

There are 5 major steps for generating four types of guidance using a local rule-based model-agnostic technique.
First, for each instance to be explained ($i_{\mathrm{explain}}$), we select the nearest instances surrounding such an instance to be explained from the training set ($I_{\mathrm{nearest}}$), \emph{cf.} Line 1.
Second, we generate synthetic instances ($I_{\mathrm{synthetic}}$) around the neighbourhood of each instance to be explained, \emph{cf.} Line 2.
Then, we create a set of combined instances  as $I_{\mathrm{combined}} = I_{\mathrm{nearest}} \cup I_{\mathrm{synthetic}}$, \emph{cf.} Line 3, which is a combination of the nearest instances and the synthetic instances.
Third, we use the global defect prediction models to generate the predictions of the combined instances (i.e., $P_{I_{\mathrm{combined}}}$), \emph{cf.} Line 4.
Fourth, to learn the associations between the synthetic features and the predictions of the global defect prediction models, we use the Magnum Opus association rule learning algorithm~\cite{webb1995opus} to generate a set of optimal association rules that are the most predictive (i.e., rules with the highest confidence) and the most interesting (i.e., rules with the highest lift) from the combined instances and their predictions, \emph{cf.} Line 5.
Finally, we classify the set of association rules into four types of rule-based guidance with respect to a contingency table of such association rules and identify the best rule for each type of guidance, \emph{cf.} Line 6.
Below, we explain each major step in details.


\begin{figure}[t]
	\includegraphics[width=\columnwidth]{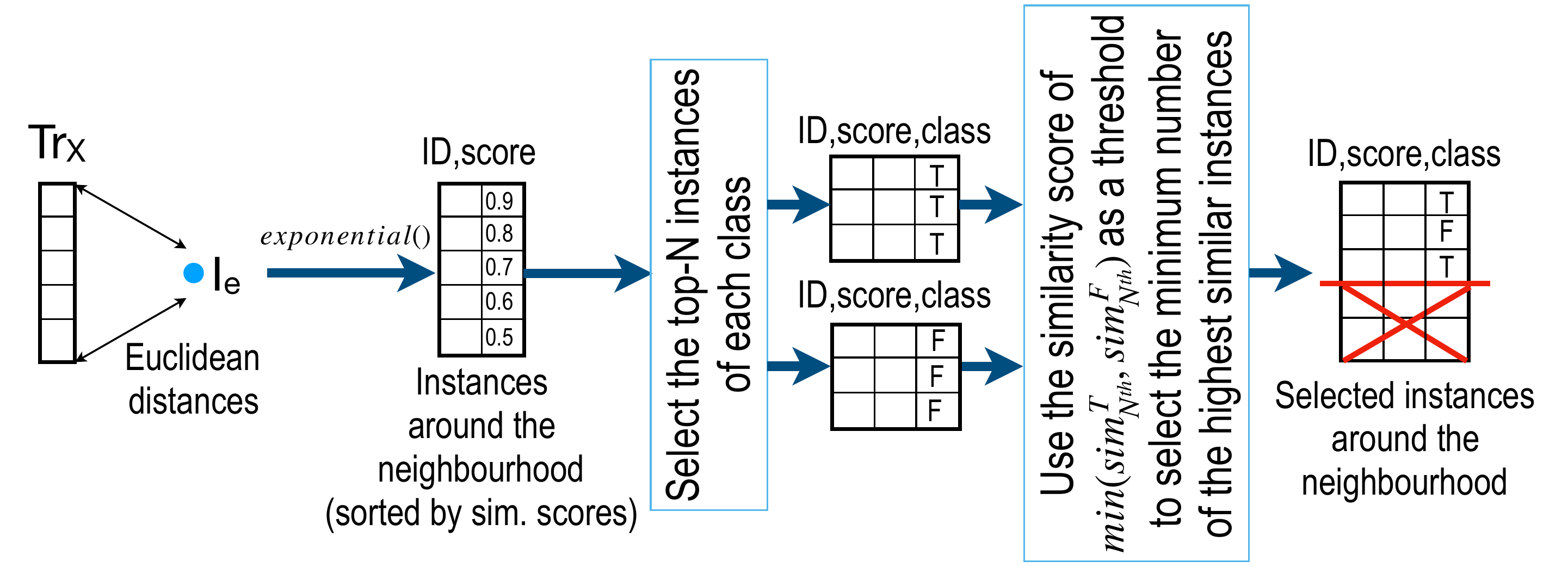}
	\caption{An approach to select instances around the neighbourhood.}
	\label{fig:selected_instances}
\end{figure}

\subsubsection*{Phase 2-1: Select the nearest instances surrounding an instance to be explained}\label{sec:sampling}


We assume that instances from the neighbourhood of the instance to be explained have approximately equivalent characteristics to an instance to be explained. 
Figure~\ref{fig:selected_instances} presents an overview of the steps to select the nearest instances from the neighbourhood of the instance to be explained. 
In particular, there are three steps as follows:

\textbf{(Step 1) -- Normalize feature values.} 
Different features may have different units and thus their range values may vary greatly.
For example, $\mathrm{LOC}$ (e.g., 100 lines of code) and $\mathrm{Ownership}$ (e.g., an ownership score of 0.5).
Thus, we first apply a Z-score normalization to each feature in defect datasets.

\textbf{(Step 2) -- Compute the similarity scores of instances in training data.}
To do so, we first compute the Euclidean distance between the instances in the training data ($Tr_{x}$) and the instance to be explained ($i_e$). 
Then, we apply an exponential kernel function to convert such Euclidean distances into similarity scores.
using an exponential kernel function to make the distance more linearly distributed.
%




\textbf{(Step 3) -- Select the smallest number of the most similar instances using the top-N instances of each class.} 
To do so, we first sort the similarity scores of instances ($sim$) in descending order for each class.
Then, we select the top $N$ instances of each class from the sorted similarity scores.
The lowest similarity score of the top $N$ instances of each class (i.e., $\mathrm{Min}(sim^{\mathrm{True}}_{N^{\mathrm{th}}}, sim^{\mathrm{False}}_{N^{\mathrm{th}}})$) is used as a threshold to select the minimum number of the most similar instances.
Such the lowest similarity score among the top $N$ instances of both classes is used to determine the boundary of the neighbourhood.
For example, given an example of $N=10$, the lowest similarity scores of the top-10 instances with the highest similarity scores of $\mathrm{DEFECT}$ and $\mathrm{CLEAN}$ classes are 0.8 and 0.9, respectively.
Therefore, in this example, the similarity score of 0.8 (the 10th instance from class $\mathrm{DEFECT}$) is used to determine the boundary of the neighbourhood.
The selected instances are instances that have the similarity scores of above 0.8 (i.e., $sim \geq 0.8$).

\begin{algorithm}[tb]
	\caption{A Local Rule-based Model Interpretability with $k$-optimal Associations}\label{Alg:lormika}
	\footnotesize
	\DontPrintSemicolon
	\SetKwInOut{Input}{Input}
	\SetKwInOut{Output}{Output}
	\Input{{\jingAlgo{Tr_x}} {\text{training instances  without target}}}
	\nonl\myinput{~~{\jingAlgo{Tr_y}} {\text{target (class label) of training instances}}}
	\nonl\myinput{~~{\jingAlgo{i_{\mathrm{explain}}}} {\text{an instance need to be explained}}}
	\nonl\myinput{~~{\jingAlgo{M}} {\text{ a global defect prediction model}}}
	\nonl\myinput{~~{\jingAlgo{N_\mathrm{features}}} {\text{\# of features}}}
	\nonl\myinput{~~{\jingAlgo{N_\mathrm{synthetic}}} {\text{\# of the new instances to be generated}}}
	\Output{$G_{i_{\mathrm{explain}}} ~- \text{Four types}$ of rule-based guidance for the \newline instance to explain $i_{\mathrm{explain}}$}
	\BlankLine
	\begin{algorithmic}[1]
	\SetAlgoLined
		\State{\outAlgo{I_{\mathrm{nearest}}} $\text{SelectFromNeighbourhood}(Tr_{px},i_{\mathrm{explain}})$}
		\State{\outAlgo{I_{\mathrm{synthetic}}}$\text{GenerateFromNeighbourhood}(I_{\mathrm{selected}}, \newline \hspace{2cm}~~~~~~~~~~~~~~~~~N_\mathrm{features},N_\mathrm{synthetic},i_{\mathrm{explain}})$}
		\State{\outAlgo{I_{\mathrm{combined}}}$ I_{\mathrm{nearest}} \cup I_{\mathrm{synthetic}} $}
		\State{\outAlgo{P_{I_{\mathrm{combined}}}} $\text{GetPredictFromGlobalModel}(I_{\mathrm{combined}},M)$}
	\State{\outAlgo{R_{i_{\mathrm{explain}}}} $\text{GenerateMagnumOpusRules}(I_{\mathrm{combined}}, P_{I_{\mathrm{combined}}})$}
		\State{\outAlgo{G_{i_{\mathrm{explain}}}} $\text{GenerateRuleGuidance}(R_{i_{\mathrm{explain}}}, i_{\mathrm{explain}},P_{i_{\mathrm{explain}}})$}
		\State \Return $G_{i_{\mathrm{explain}}}$		
\end{algorithmic}
\end{algorithm}

\subsubsection*{Phase 2-2: Generate synthetic instances to expand the neighbourhood}\label{sec:generating}

The number of selected nearest instances in the neighbourhood may not be enough to accurately learn the behaviour of the instance to be explained.
Thus, we generate synthetic instances to expand the neighbourhood. 
To do so, we use the crossover (or interpolation) technique and the mutation technique to generate new synthetic instances while ensuring that the majority of such synthetic instances are within the neighbourhood of the instance to be explained.
Below, we describe how we generate synthetic instances using the crossover and the mutation techniques in details.


\textbf{Generate synthetic instances using the crossover technique.} 
To do so, we randomly select two different instances from the neighbourhood of the instance to be explained.
Then, we generate the synthetic instances based on the crossover technique using the following equation:
\begin{equation} \label{Eq:Interpolation_Equation}
I_{\mathrm{crossover}} = x + (y-x)*\alpha
\end{equation}

\noindent where $x$ and $y$ are random parent instances from the training set, and $\alpha$ is a randomly generated number between $0$ and $1$. 

\textbf{Generate synthetic instances using the mutation techniques.} 
To do so, we randomly select three different instances from the neighbourhood of the instance to be explained. 
Then, we generate synthetic instances based on the mutation technique~\cite{Storn1997-wv} using the following equation:
\begin{equation} \label{Eq:mutation_Equation}
I_{\mathrm{mutation}} = x + (y-z)*\mu
\end{equation}

\noindent where $x, y$ and $z$ are random parent instances from the training set, and $\mu$ is a randomly generated number between $0.5$ and $1$.




\subsubsection*{Phase 2-3: Generate the predictions of the nearest instances and the synthetic instances from defect prediction models}

Firstly, we name a set of such the nearest instances (generated in Phase 2-1) and the synthetic instances (generated in Phase 2-2) as the combined instances $I_{\mathrm{combined}}$, where $I_{\mathrm{combined}} = I_{\mathrm{nearest}} \cup I_{\mathrm{synthetic}}$
Then, we generate the predictions of such combined instances in the neighbourhood (i.e., $\mathrm{Prediction}_{I_{\mathrm{nearest}} \cup I_{\mathrm{synthetic}}}$) from defect prediction models to learn the behaviour and the logics of such defect prediction models.

\subsubsection*{Phase 2-4: Generate association rules using Magnum Opus association rule mining}
\label{sec:associationrules}

\revised{R1.9}{
The Magnum OPUS association rule mining algorithm performs statistically sound association rule mining by combining $k$-optimal association discovery techniques~\cite{webb2005k} and the OPUS search algorithm~\cite{webb1995opus} to find the $k$ most interesting associations according to a defined criterion (e.g., lift, confidence, coverage). 
The effectiveness of our SQAPlanner relies on this algorithm to generate the rule-based explanations. 
With the functionality of the OPUS search algorithm, it will effectively prune the search space by discarding the associations which are likely to be spurious, and removing false positives by performing Fisher's exact hypothesis test.
}
We use an implementation of the $k$-optimal association rule mining technique as provided by the BigML platform.\footnote{https://bigml.com/}

\subsubsection*{Phase 2-5: Generate four types of rule-based guidance}
\label{sec:rulesbg}

Finally, we classify the optimal set of association rules that are identified by Magnum OPUS into four categories with respect to a contingency table of the LHS and RHS of the association rules.
Then, we identify the best rule that is the most predictive and the most interesting for each type of guidance as the output of SQAPlanner. 

To better illustrate how we classify the output rules generated by Magnum OPUS, we use four examples of an association rule as a subject of this explanation.
Given an instance to explain $i_{\mathrm{explain}}$ that has 200 lines of code ($\mathrm{LOC}=200$) and is predicted as $\mathrm{DEFECT}$ by the global defect prediction model, our SQAPlanner framework generates the following four types of rule-based explanations:

\begin{enumerate}[{\bf G1:}]
\item \textbf{Risky current practices that lead the defect model to predict a file as defective.} \\
\emph{Technical Name.} Supporting Rules ($\Re^{+}$).\\
\emph{Definition.} if LHS = true, then RHS = true.\\
\emph{Example.} $\{ \mathrm{LOC} > 150\} \xLongrightarrow{\text{associate}} \mathrm{DEFECT}$ \\
\emph{Interpretation.} This example is a supporting rule, since (1) the antecedent (LHS) of the rule hold true as the actual $\mathrm{LOC}$ of $i_{\mathrm{explain}}$ (i.e., 200) is actually higher than 150, and (2) the consequent (RHS) of the rule hold true as the prediction of $i_{\mathrm{explain}}$ generated by the global defect prediction model is $\mathrm{DEFECT}$.


\item \textbf{Non-risky current practices that lead the defect model to predict a file as clean.} \\
\emph{Technical Name.} Contradicting Rules ($\Re^{-}$).\\
\emph{Definition.} if LHS = true, then RHS = false. \\
\emph{Example.} $\{ \mathrm{LOC} < 500\} \xLongrightarrow{\text{associate}} \mathrm{CLEAN}$ \\
\emph{Interpretation.} This example is a contradicting rule, since (1) the antecedent (LHS) of the rule hold true as the actual $\mathrm{LOC}$ of $i_{\mathrm{explain}}$ (i.e., 200) is actually lower than 500, yet (2) the consequent (RHS) of the rule does not hold true as the prediction of $i_{\mathrm{explain}}$ generated by the global defect prediction model is $\mathrm{DEFECT}$.

\item \textbf{Potential practices to avoid to not increase the risk of having defects.} \\
\emph{Technical Name.} Hypothetical Supporting Rules ($\Re^{H+}$).\\
\emph{Definition.} if LHS = false, then RHS = true.\\
\emph{Example.} $\{ \mathrm{LOC} > 300\} \xLongrightarrow{\text{associate}} \mathrm{DEFECT}$ \\
\emph{Interpretation.} This example is a hypothetical supporting rule, since (1) the antecedent (LHS) of the rule does not hold true as the actual $\mathrm{LOC}$ of $i_{\mathrm{explain}}$ (i.e., 200) is not higher than 300, yet (2) the consequent (RHS) of the rule hold true as the prediction of $i_{\mathrm{explain}}$ generated by the global defect prediction model is $\mathrm{DEFECT}$.

\item \textbf{Potential practices to follow to decrease the risk of having defects.}  \\
\emph{Technical Name.} Hypothetical Contradicting Rules or Counterfactual Rules ($\Re^{H-}$).\\
\emph{Definition.} if LHS = false, then RHS = false.\\
\emph{Example.} $\{ \mathrm{LOC} <100 \} \xLongrightarrow{\text{associate}} \mathrm{CLEAN}$ \\
\emph{Interpretation.} This example is a hypothetical contradicting rule, since (1) the antecedent (LHS) of the rule does not hold true as the actual $\mathrm{LOC}$ of $i_{\mathrm{explain}}$ (i.e., 200) is not lower than 100, and (2) the consequent (RHS) of the rule does not hold true as the prediction of $i_{\mathrm{explain}}$ generated by the global defect prediction model is $\mathrm{DEFECT}$.
\end{enumerate}

\begin{table*}[t]
\caption{A statistical summary of the studied systems.}
\label{Table:StudiedSystems}
\centering
\resizebox{\textwidth}{!}{
\begin{tabular}{lllllll} 
\hline
Name & Description & \#DefectReports & No. of files & Defective Rate & KLOC & Studied Releases \\
\hline
ActiveMQ & Messaging and Integration Patterns server & 3,157 & 1,884-3,420 & 6\%-15\% & 142-299 &5.0.0, 5.1.0, 5.2.0, 5.3.0, 5.8.0 \\
Camel & Enterprise Integration Framework & 2,312 & 1,515-8,846 & 2\%-18\% & 75-383 & 1.4.0, 2.9.0, 2.10.0, 2.11.0 \\
Derby & Relational Database & 3,731 & 1,963-2,705 & 14\%-33\% & 412-533 & 10.2.1.6, 10.3.1.4, 10.5.1.1\\
Groovy & Java-syntax-compatible OOP for JAVA & 3,943 & 757-884 & 3\%-8\% & 74-90 & 1.5.7, 1.6.0.Beta\_1, 1.6.0.Beta\_2\\
HBase & Distributed Scalable Data Store & 5,360 & 1,059-1,834 & 20\%-26\% & 246-534 &  0.94.0, 0.95.0, 0.95.2 \\
Hive & Data Warehouse System for Hadoop & 3,306 & 1,416-2,662 & 8\%-19\% & 287-563 & 0.9.0, 0.10.0, 0.12.0 \\
JRuby & Ruby Programming Lang for JVM & 5,475 & 731-1,614 & 5\%-18\% & 105-238 & 1.1, 1.4, 1.5, 1.7\\ 
Lucene & Text Search Engine Library & 2,316 & 8,05-2,806 & 3\%-24\% & 101-342 & 2.3.0, 2.9.0, 3.0.0, 3.1.0 \\
Wicket & Web Application Framework & 3,327 & 1,672-2,578 & 4\%-7\% & 109-165 & 1.3.0.beta1, 1.3.0.beta2, 1.5.3\\
\hline
\end{tabular}}
\end{table*}

\renewcommand{\arraystretch}{1.2}
\begin{table*}
\caption{A summary of the studied code metrics.}
\label{tab:staticmetrics}
 \centering
 \begin{tabular}{p{0.065\linewidth}p{0.835\linewidth}p{0.0325\linewidth}}
 \hline
 \textbf{Granularity} & \textbf{Metrics} & \textbf{Count}\\
 \hline
 File & AvgCyclomatic, AvgCyclomaticModified, AvgCyclomaticStrict, AvgEssential, AvgLine, AvgLineBlank, AvgLineCode, AvgLineComment, CountDeclClass, CountDeclClassMethod, CountDeclClassVariable, CountDeclFunction, CountDeclInstanceMethod, CountDeclInstanceVariable, CountDeclMethod, CountDeclMethodDefault, CountDeclMethodPrivate, CountDeclMethodProtected, CountDeclMethodPublic, CountLine, CountLineBlank, CountLineCode, CountLineCodeDecl, CountLineCodeExe, CountLineComment, CountSemicolon, CountStmt, CountStmtDecl, CountStmtExe, MaxCyclomatic, MaxCyclomaticModified, MaxCyclomaticStrict, RatioCommentToCode, SumCyclomatic, SumCyclomaticModified, SumCyclomaticStrict, SumEssential & 37\\
 \cline{1-3}
 Class & CountClassBase, CountClassCoupled, CountClassDerived, MaxInheritanceTree, PercentLackOfCohesion & 5\\
 \cline{1-3}
 Method & CountInput\_\{Min, Mean, Max\}, CountOutput\_\{Min, Mean, Max\}, CountPath\_\{Min, Mean, Max\}, MaxNesting\_\{Min, Mean, Max\} & 12\\
 \hline
 \end{tabular}
\hspace{\fill}
\end{table*}

\renewcommand{\arraystretch}{1.2}
\begin{table}
\caption{A summary of the studied process and ownership metrics.}
\label{tab:processmetrics}
\resizebox{.99\columnwidth}{!}{ 
\centering
\begin{tabular}{p{0.23\linewidth}p{0.8\linewidth}}
\hline
\textbf{Metrics} & \textbf{Description}\\
\hline
\multicolumn{2}{l}{\emph{Process Metrics}}    \\
\hline
COMM & The number of Git commits\\
ADDED\_LINES & The normalized number of lines added to the module \\
DEL\_LINES & The normalized number of lines deleted from the module \\
ADEV & The number of active developers\\
DDEV & The number of distinct developers\\
\hline
\multicolumn{2}{l}{\emph{Ownership Metrics}}    \\
\hline
MINOR\_COMMIT & The number of unique developers who have contributed less than 5\% of the total code changes (i.e., Git commits) on the module\\
MINOR\_LINE & The number of unique developers who have contributed less than 5\% of the total lines of code on the module\\ 
MAJOR\_COMMIT & The number of unique developers who have contributed more than 5\% of the total code changes (i.e., Git commits) on the module \\
MAJOR\_LINE & The number of unique developers who have contributed more than 5\% of the total lines of code on the module\\
OWN\_COMMIT & The proportion of code changes (i.e., Git commits) made by the developer who has the highest contribution of code changes on the module\\ 
OWN\_LINE & The proportion of lines of code written by the developer who has the highest contribution of lines of code on the module \\
\hline
\end{tabular}}
\end{table}

\section{Experimental Design and Results}\label{sec:results}

In this section, we aim to investigate (RQ3) the effectiveness,
(RQ4) the stability,  
and (RQ5) the applicability of the rule-based explanations generated by our SQAPlanner.  
Below, we describe the studied projects, the experimental design, and present the results.


\vspace{-4mm}
\subsection{Studied Projects}\label{sec:projects}

To select some suitable projects, we identified three important criteria that need to be satisfied:

\begin{itemize}
\item \textbf{Criterion 1 --- Publicly-available defect datasets}:
  To support verifiability and foster replicability of our study, we choose to train our defect prediction models using publicly available defect datasets.

\item \textbf{Criterion 2 --- Multiple releases}:
  The central hypothesis of our approach is that the guidance that is derived from past knowledge (a release $k-1$) can be used to explain the predictions of defective files in the target releases (a release $k$) and be applicable to prevent software defects in future releases (a release $k+1$).
  Thus, we need multiple releases for each software project to validate our hypothesis.

\item \textbf{Criterion 3 --- Labels of defective files are based on actual affected releases}:
  Prior work raises concerns that the approximation of the post-release window periods (e.g., 6 months) that are popularly-used in many defect datasets may introduce bias to the construct to the validity of our results~\cite{yathish2019affectedrelease}. Instead of relying on traditional post-release window periods, we choose to use defect datasets that are labeled based affected releases, as suggested by recent studies~\cite{yathish2019affectedrelease,da2017framework}. 
\end{itemize}

Thus, we finally selected a corpus of publicly available defect datasets provided by Yatish~\ea~\cite{yathish2019affectedrelease} where the ground-truths are labeled based on the affected releases. 
These datasets consist of 32 releases that span 9 open-source, real-world, non-trivial software systems. 
Table~\ref{Table:StudiedSystems} shows a statistical summary of the studied datasets.
Each dataset has 65 software metrics along 3 dimensions, i.e., 54 code metrics, 5 process metrics, and 6 human metrics. 
Table~\ref{tab:staticmetrics} shows a summary of the static code metrics, while Table~\ref{tab:processmetrics} shows a summary of the process and human metrics.
The full details of the data collection process are available at Yatish~\ea~\cite{yathish2019affectedrelease}.

\subsection{Experimental Design}

\begin{figure}[t]
	\centering
	\includegraphics[width=\columnwidth]{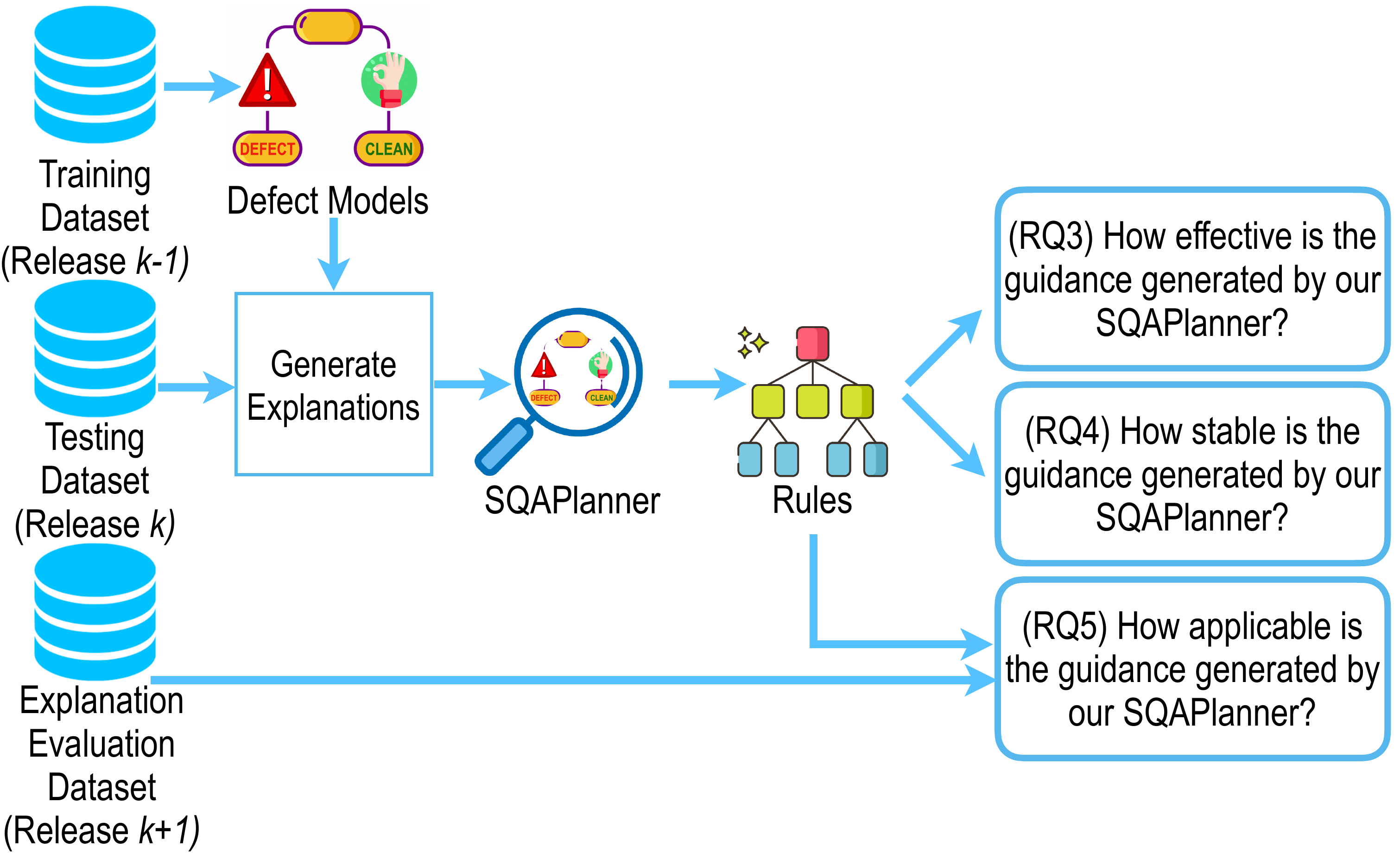}
	\caption{An evaluation framework of our SQAPlanner approach}
	\label{fig:evalution}
\end{figure}

\revised{R2.17}{
We hypothesize that the guidance that is derived from past knowledge (a release $k-1$) can be used to explain the predictions of defective files in the target releases (a release $k$) and be applicable to prevent software defects in future releases (a release $k+1$).
Thus, we evaluate our approach (see Figure~\ref{fig:evalution}) using a set of three consecutive releases ($k$-1, $k$, and $k$+1) for training, testing, and explanation evaluation, respectively.
We first trained our defect models using a random forest classification technique on a training release (i.e., a release $k-1$).
Then, we generate rule-based explanations for each file in the testing release (i.e., a release $k$).
Finally, we evaluate the applicability of the rule-based explanations with the explanation evaluation release (i.e., a release $k+1$).
Let's take an example of the ActiveMQ project, we first use the release 5.0.0 for training, the release 5.1.0 for testing, and the release 5.2.0 for explanation evaluation. 
We repeat the experiment similarly for the other consecutive releases (i.e., \{5.1.0, 5.2.0, 5.3.0\}, \{5.2.0, 5.3.0, 5.8.0\}) and for other projects.
}

\subsection{Results}

\subsection*{(RQ3) \rqiii}

\smallsection{Motivation}
Our SQAPlanner is based on the assumption that our rule-based explanations are generated based on the approximation of the characteristics of files that are similar to the file to be explained.
This assumption is similar to those of many local rule-based model-agnostic techniques~\cite{ribeiro2016should,guidotti2018local,ribeiro2018anchors} that the behaviour of the instance to be explained is similar to the behaviours of the instances around its neighbourhood.
According to the definition of rule-based explanations in Section~\ref{sec:guidance}, our SQAPlanner generated rule-based explanations will be considered effective if such rule-based explanations achieve a high coverage and high confidence value.


\smallsection{Approach} To address RQ3, we evaluate the rule-based explanations generated by our SQAPlanner using the traditional association rule evaluation measures (i.e., coverage, confidence, and lift).

\textbf{Coverage} measures support of the antecedent of an association rule, i.e., the percentage of files that support the rule conditions.
Formally,
$\text{Coverage}(p~\rightarrow~q) = \text{Support}(p)$ 
where Support($p$) is the proportion of files that fulfill $p$. 

\begin{center}
$\text{Support}(p) = \dfrac{|\text{files} \in \text{Dataset, such that files fulfill p}| }{\text{\#total files}}$
\end{center}

For example, a rule-based explanation (G1) of
$\{ \mathrm{DEV} > 10\} \xLongrightarrow{\text{associate}} \mathrm{DEFECT}$ with a coverage value of 0.9 indicates that 90\% of the files fulfill a risky practice of having more than ten developers who touch a file.
A high coverage value of the G1 guidance indicates that such a risky practice is a common risky practice to many files of the dataset.

\textbf{Confidence (i.e., Precision or Strength)} measures the percentage of files that fulfill the antecedent and consequent together over the number of files that only fulfill the antecedent, which can be defined as follows: 

$\text{Confidence}(p \rightarrow q) = \text{Support}(p \rightarrow q)/\text{Support}(p)$.
\\
For example, a rule-based explanation (G1) of $\{ \mathrm{DEV} > 10\} \xLongrightarrow{\text{associate}} \mathrm{DEFECT}$ with a confidence value of 0.8 indicates that, there are 80\% of the files that fulfill the risky practice of having more than ten developers who touch a file are actually defectives.
A high confidence value of the G1 guidance indicates that such risky practice is a high confident risky practice to many files of the dataset.

\textbf{Lift} measures how many times more often the antecedent and consequent occur together compared to what would be expected when they (i.e., both antecedent and consequent) were statistically independent, which can be defined as follows:
\begin{center}
$\text{Lift}(p \rightarrow q) = \frac{\text{Support}(p \rightarrow q)}{\text{Support}(p) \times \text{Support}(q)} \nonumber$
\end{center}

For example, a rule-based explanation (G1) of $\{ \mathrm{DEV} > 10\} \xLongrightarrow{\text{associate}} \mathrm{DEFECT}$ with a life value of 5 indicates that, the file will be 5 times (i.e., 500\%) more likely to be defective if the rule is fulfilled.
A lift value greater than one means that a file is likely to be defective if the conditions are fulfilled, while a lift value less than one means a file is unlikely to be defective if the conditions are fulfilled.
A high lift value of the G1 guidance indicates that there is a high chance that a file is likely to be defective if such risky practice is fulfilled.
Thus, practitioners should pay attention to guidance rules with a high lift value.

\textbf{Baseline comparison.} We compare our SQAPlanner with the two state-of-the-art local rule-based model-agnostic techniques, i.e.,  Anchor~\cite{ribeiro2018anchors} and LORE~\cite{guidotti2018local}~\cite{ribeiro2016should}.

\textit{Anchor}, an extension of LIME~\cite{ribeiro2016should}, was proposed by Ribeiro~\ea~~\cite{ribeiro2018anchors}. 
The key idea of Anchor is to select \emph{if-then} rules -- so-called \emph{anchors} -- that have high confidence, in a way that features that are not included in the rules do not affect the prediction outcome if their feature values are changed.
In particular, Anchor selects only rules with a minimum confidence of 95\%, and then selects the rule with the highest coverage if multiple rules have the same confidence value.

\textit{LORE} is proposed by Guidotti~\ea~\cite{guidotti2018local}.
For each instance to be explained, LORE generates files around the neighbourhood using a genetic algorithm. 
LORE then obtains predictions of the generated files from the global defect models to learn the behaviour and the logics of the defect models. Finally, a decision tree is built on the defined neighbourhood of the instance to be explained and is then later converted to rules.


 \begin{figure}[t]
     \centering
     \includegraphics[width=\columnwidth]{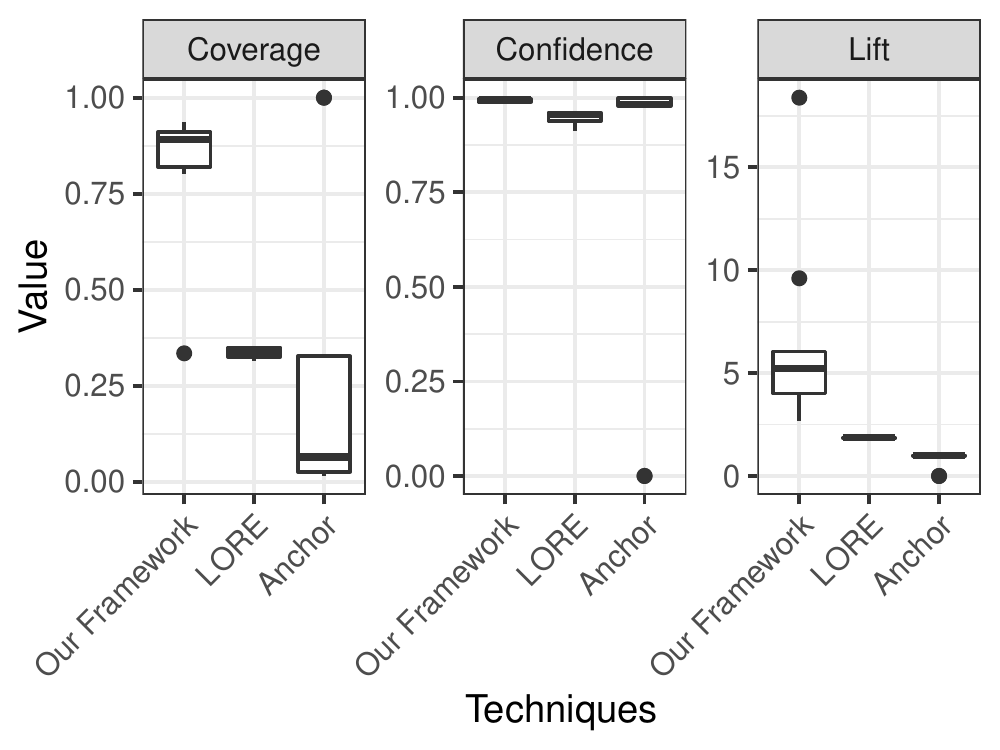}
     \caption{(RQ3) The distribution of the evaluation measures of our rule-based explanations when compared to baseline approaches (i.e., LORE and Anchor).}
     \label{Fig:coverage}
 \end{figure}
 
\smallsection{Results} Figure~\ref{Fig:coverage} presents the results for coverage, confidence, and lift of the local rule-based model-agnostic techniques. 

\textbf{(Coverage) At the median, 89\% of files are supported by the rule-based explanations, suggesting that our SQAPlanner outperforms the LORE and Anchor local rule-based model-agnostic techniques.}
Figure~\ref{Fig:coverage} shows that the median coverage is 89\%, 34\%, and 6\% for our SQAPlanner, LORE, and Anchor, respectively.
We suspect that the high coverage values that are achieved by our SQAPlanner are due to the flexibility of the $k$-optimal search that allows us to search particularly for rules with high coverage. 
High coverage is important as it is a measure for how representative a rule is for a given dataset, so that our results suggest that our SQAPlanner achieves the most representative rules.


\textbf{(Confidence) At the median, 99\% of files are supported by the antecedent and the consequent of the rule-based explanations, which outperforms the LORE and Anchor model-agnostic techniques.}
Figure~\ref{Fig:coverage} shows the distributions of the confidence for our SQAPlanner, LORE, and Anchor, respectively.
We find that LORE and Anchor achieve high confidence with median confidence of 95\% and 98\%, respectively.
We find that the comparable confidence values achieved by LORE and Anchor have to do with the main optimization goal of Anchor and LORE, since both LORE and Anchor techniques aim to search for rules with the highest confidence. 
Nevertheless, we find that our SQAPlanner achieves the highest median confidence of 99\%.

 
\textbf{(Lift) The rule-based explanations generated by our SQAPlanner achieve a median lift value of 6.6, which outperforms the LORE and Anchor model-agnostic techniques.}
Figure~\ref{Fig:coverage} shows that the median lift is 6.6, 5.2, 0.98 for our SQAPlanner, LORE and Anchor respectively.
The highest lift value of 6.6 indicates that files will be 6.6 times (i.e., 660\%) more likely to be defective if the rule is matched.
Similarly, the highest lift value of our SQAPlanner can be attributed to the flexibility of the $k$-optimal search that allows us to search particularly for rules with the highest lift.
On the other hand, Anchor achieves a lower lift score, since Anchor constructs the neighbourhood in a way that it contains only files of the same class as the instance in consideration. 
Thus, the lift scores for Anchor under these circumstances are equal to the confidence values.
\subsection*{(RQ4) \rqiv}

\smallsection{Motivation} Our SQAPlanner approach and the two state-of-the-art local rule-based model-agnostic techniques (i.e., LORE and Anchor) involve random data generation when generating synthetic instances around the neighbourhood.
As such, the randomization bias may produce different rule-based explanations when the approaches are re-executed.
Thus, we aim to investigate the consistency of the rule-based explanation of the same instance when these model-agnostic techniques are re-executed.

\smallsection{Approach} To address RQ4, we repeat our experiment ten times to investigate the stability of our rules.
Since the rules generated by the baseline comparison are optimized based on confidence only, we focus on the rules generated by our approach that are optimized for confidence as well.

For each rule-based explanation of each file, we use the Jaccard coefficient to measure the consistency of the generated rule-based explanations.
The Jaccard coefficient compares the common and the distinct features in two given sets (e.g., $X$ and $Y$) using the following equation: $J(X,Y)=|X \cap Y|/|X\cup Y|$.
The coefficient ranges from 0\% to 100\%. 
The higher the coefficient the higher the similarity of rules over two independent runs. 

\begin{figure}[t]
\centering
\includegraphics[width=.7\columnwidth]{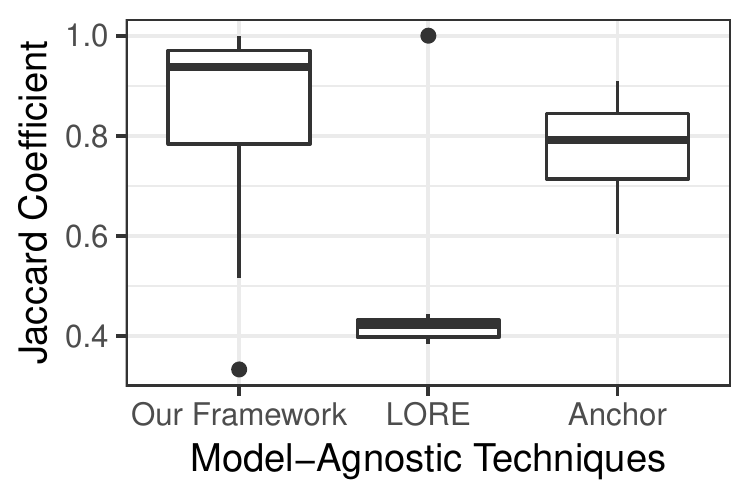}
\caption{(RQ4) The distribution of the Jaccard Coefficients of the rule-based model-agnostic techniques.}
\label{Fig:jaccard}
\end{figure}

\smallsection{Results} \textbf{Our SQAPlanner approach produces the most consistent rule-based explanations when compared to LORE and Anchor.} 
Figure~\ref{Fig:jaccard} shows that our SQAPlanner achieves a median Jaccard coefficient of 0.92, while LORE and Anchor achieve a median Jaccard coefficient of 0.42, and 0.79, respectively.
In other words, for each prediction of an instance to be explained, our rule-based explanations are (at the median) 92\% consistent with the rule-based explanations when re-executing our framework in multiple independent runs. 
In addition, our SQAPlanner's rule-based explanations are (at the median) 13\% and 50\% more consistent than the rule-based explanations generated by Anchor and LORE, respectively.
We suspect that the highest consistency achieved by our approach is a result of the more robust nature of our framework when selecting similar instances from the training data and when generating synthetic instances around the neighbourhood (as described in Sections~\ref{sec:sampling} and \ref{sec:generating}).
In contrast, Anchor uses a bandit algorithm~\cite{kocsis2006bandit} to generate neighbours, while LORE uses a genetic algorithm to generate neighbours.


\subsection*{(RQ5) \rqv}

\begin{figure*}[t]
\centering
\includegraphics[width=.95\textwidth]{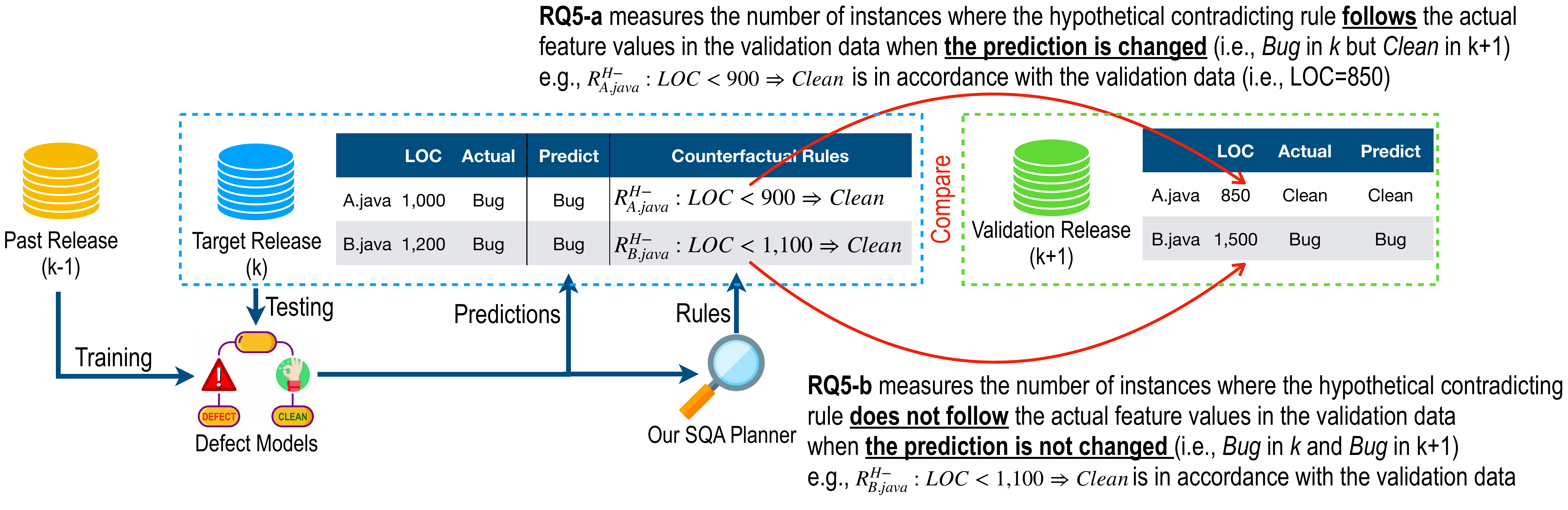}
\caption{(RQ5) An approach to evaluate the applicability of the hypothetical contradicting rules.}
\label{fig:rq3-approach}
\end{figure*}

\smallsection{Motivation} The central hypothesis of our approach is that the rule-based explanations derived from past knowledge (a release $k-1$) can be used to explain the predictions of defective files in a target release (a release $k$), and thus be applicable to guide SQA planning to prevent software defects in future releases (release $k+1$).
We want to investigate what are the proportion of files where the rule-based explanations are satisfied and not satisfied with the actual feature values in the subsequent release.

\smallsection{Approach} To address RQ5, we focus on the hypothetical contradicting rules, which are rules that guide what are the practices to follow to decrease the risk of having defects (i.e, whether the prediction of the same file could be reversed if the rule is followed in a subsequent release).
We note that not all of the files whose hypothetical contradicting rules can be generated by Anchor and LORE, since we find that LORE produces a maximum of 69\% hypothetical contradicting rules across projects (median amount of rules produced per project is 41\%), and Anchor by definition does not generate any hypothetical contradicting rules.
Since our approach is the only one that can generate hypothetical contradicting rules, we focus only on our SQAPlanner approach.
Figure~\ref{fig:rq3-approach} presents an approach to evaluate the applicability of the hypothetical contradicting rules. 
We analyze the applicability of the hypothetical contradicting rules along 2 perspectives:

\textbf{RQ5-a: Are hypothetical contradicting rules \underline{applied} when the prediction of an instance \underline{changes} from defective in a testing release $k$ to clean in a validation release $k+1$?}
Hypothetical contradicting rules are considered as applicable if such rules follow the actual feature values in the validation data when the prediction of the instance changes from defective in $k$ to clean in $k+1$.
For example, A.java is predicted to be defective in the testing data ($k$) but predicted to be clean in the validation data ($k+1$).
We consider that the generated hypothetical contradicting rules (e.g., $\{ \mathrm{LOC} < 900\} \xLongrightarrow{\text{associate}} \mathrm{CLEAN}$) is correct if such rule is in accordance with the actual feature values in the validation data (i.e., $\mathrm{LOC}$ = 850).
Like in this example, the hypothetical contradicting rule suggests developers reduce the lines of code to less than 900 to potentially reverse the decision of the defect models from defective to clean, which is consistent with the validation data ($\mathrm{LOC}$ = 850).

\textbf{RQ5-b: Are hypothetical contradicting rules \underline{not applied} when the prediction of an instance does \underline{not change} from defective in a testing release $k$ to clean in a validation release $k+1$?}
Hypothetical contradicting rules are considered applicable if such rules do not follow the actual feature values in the validation data when the prediction of the instance does not change from defective in $k$ to clean in $k+1$.
For example, we consider B.java to be predicted to be defective in both testing data and validation data.
Thus, we consider that the generated hypothetical contradicting rule (e.g., $\{ \mathrm{LOC} < 1100\} \xLongrightarrow{\text{associate}} \mathrm{CLEAN}$) is applicable if such rule does not follow the actual feature values in the validation data (i.e., $\mathrm{LOC}$ = 1,500).

\begin{figure}[t]
    \centering
    \includegraphics[width=\columnwidth]{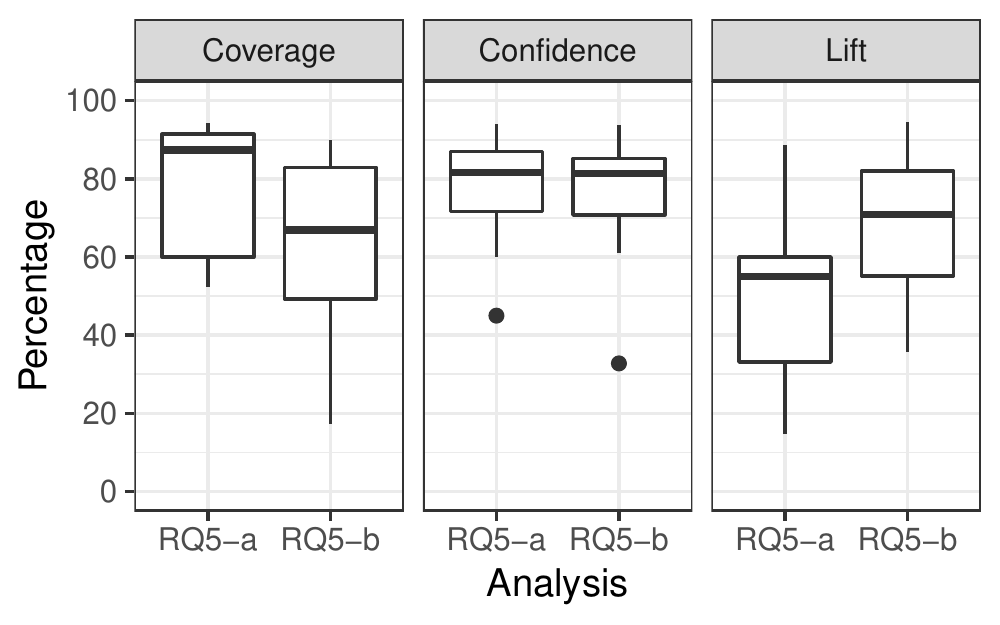}
    \caption{The percentage of of the instances in the subsequent releases where our hypothetical contradicting rules (RQ5-a) follow the actual feature values in the validation data when the decision is changed and (RQ5-b) do not follow the actual feature values in the validation data when the decision is not changed for each measure.}
    \label{Fig:rq4}
\end{figure}

For each perspective, we compute the number of instances where the hypothetical contradicting rule does follow and does not follow the actual feature values in the subsequent release in RQ5-a and RQ5-b, respectively.
Figure~\ref{Fig:rq4} presents the proportion of files where its hypothetical contradicting rule does follow (RQ5-a) and does not follow (RQ5-b) the actual feature values in the subsequent release for each measure.

\smallsection{Results} \textbf{For 55\%-87\% of the instances in the subsequent releases, our SQAPlanner's hypothetical contradicting rules are correctly applicable when the prediction of rules changes from defective to clean.}
Figure~\ref{Fig:rq4} shows that there are 87\%, 82\% and 55\% of the instances in the subsequent releases that our hypothetical contradicting rules \emph{follow} the actual feature values in the validation data with respect to coverage, confidence, and lift, respectively.
This finding indicates that our SQAPlanner's hypothetical contradicting rules learned from  past knowledge ($k-1$) to explain the predictions of instances from the target release ($k$) could potentially reverse the predictions of the same instance in the subsequent release ($k+1$) from having defects to clean.

\textbf{For 67\%-81\% of the instances in the subsequent releases, our hypothetical contradicting rules are correctly non-applicable when the prediction of rules does not change.}
Figure~\ref{Fig:rq4} shows there are 67\%, 81\% and 71\% of the instances in the subsequent releases that our SQAPlanner's hypothetical contradicting rules \emph{do not follow} the actual feature values in the validation data when the prediction of rules does not change with respect to coverage, confidence, and lift, respectively.
In other words, when files are still defective in the subsequent release, our hypothetical contradicting rules are still largely in agreement (i.e., our hypothetical contradicting rules are correctly non-applicable).

 

\subsection{Discussion \& Qualitative Analysis}\label{sec:discussion}

We conducted a qualitative analysis to illustrate the effectiveness of our guidance generated by our SQAPlanner. 
We selected the \emph{ErrorHandlerBuilderRef.java} of the release 2.9.0 of the Camel software system as the subject of this qualitative analysis.
Our SQAPlanner approach correctly predicts this file as defective with a probability score of 70\%.
Below, we discuss the implications of our rule-based explanations to guide developers on what they could follow and could avoid to decrease the risk of having defects.

\subsubsection*{\gli}
To answer this question, we use the supporting rule to generate guidance (G1) for this file as follows:

\begin{align}
\Re^+ =  \{& \mathrm{LOCDeclaration}>28.150 \;\&\; \nonumber\\
& \mathrm{DistinctDeveloper}>1.68 \;\&\; \nonumber\\ & \mathrm{Ownership}<0.85\} \xLongrightarrow{\text{associate}} \mathrm{DEFECT}  \nonumber
\end{align}

\smallsection{Implication} This supporting rule indicates that this file is being predicted as defective since it is associated with the conditions of having more than 28 lines of declarative code, more than 1.68 distinct developers, and a line-based ownership score of less than 85\%.
When comparing  this to the actual feature values of the file \{$\mathrm{LOCDeclaration}=34$, $\mathrm{DistinctDeveloper}=3$, $\mathrm{Ownership}=0.65$\}, we find that the conditions of this supporting rule are consistent with the actual feature values and the consequent is consistent with our SQA-Planner's prediction (i.e., defective).

\subsubsection*{\glii}

To answer this question, we use our contradicting rule to generate guidance (G2) for this file as follows:

\begin{align}
\Re^- =  \{& 0.440<=\mathrm{RatioCommentToCode}<=0.960\} \nonumber \\ 
& \xLongrightarrow{\text{associate}} \mathrm{CLEAN}   \nonumber
\end{align}
\smallsection{Implication} We find that our contradicting rule is consistent with the actual feature values of the file to be explained.
The actual feature values of this file is \{$\mathrm{RatioCommentToCode}=0.51$\}, meaning that 51\% of the total lines of code are comment lines (i.e., \#comments/\#LOC).
The contradicting rule ($\Re^-$) indicates that the condition that supports its prediction as not being defective is \{$0.440<=\mathrm{RatioCommentToCode}<=0.960$\}, indicating that files that have a $\mathrm{RatioCommentToCode}$ of more than 44\% but less than 96\% are likely not to be defective. Developers should thus adhere to the contradicting rule i.e., having the comment ratio for more than 44\% of the file, to not increase the risk of having defects.


\subsubsection*{\gliii}

To answer this question, we use our hypothetical supporting rule to generate guidance (G3) for this file as follows:

\begin{align}
\Re^{H+} &=  \{\mathrm{MinorCommit}>0.000\} \xLongrightarrow{\text{associate}} \mathrm{DEFECT}  \nonumber
\end{align}
\smallsection{Implication} Having more than zero minor developers will increase the risk of having defects.
The actual feature value of this file is \{$\mathrm{MinorCommit}=0$\}, meaning that this file has no minor developers (i.e., minor) who edit or change the file. 
This finding is consistent with Bird~\ea~\cite{Bird2011a} and Rahman~\cite{rahman2011ownership} who found that minor developers often introduce defects.
Thus, developers should adhere to the hypothetical supporting rule in order not to increase the risk of having defects.

\subsubsection*{\gliv}

To answer this question, we use our hypothetical contradicting rule to generate guidance (G4) for this file as follows:

\begin{align}
\begin{split}
\Re^{H-} =  \{&\mathrm{LOCBlank}<7.62 \;\&\; \\
& \mathrm{OutputMean}<1.98\} \xLongrightarrow{\text{associate}} \mathrm{CLEAN}  \notag \nonumber
\end{split}
\end{align}
\smallsection{Implication} If developers changed the file to have less than 8 blank lines and less than 2 output variables, this could reverse the prediction of having defects to being clean.
The actual feature values of this file are \{$\mathrm{LOCBlank}=19, \mathrm{OutputMean}=3.72$\}, meaning that this file has 19 blank lines and an average of 3.7 output variables (i.e., fan-out) of functions in a file.
The hypothetical contradicting rule indicates that if \{$\mathrm{LOCBlank}<7.62 \;\&\; \mathrm{OutputMean}<1.98 $\} then the file is likely to reverse the prediction of having defects to being clean.
Thus, our hypothetical contradicting rule provides suggestions to the developers of what they should do to decrease the risk of having defects. 
It should be noted that our contradicting rule shows correlations that may not necessarily be causal.

\begin{figure*}[t]
\centering
\includegraphics[width=\textwidth]{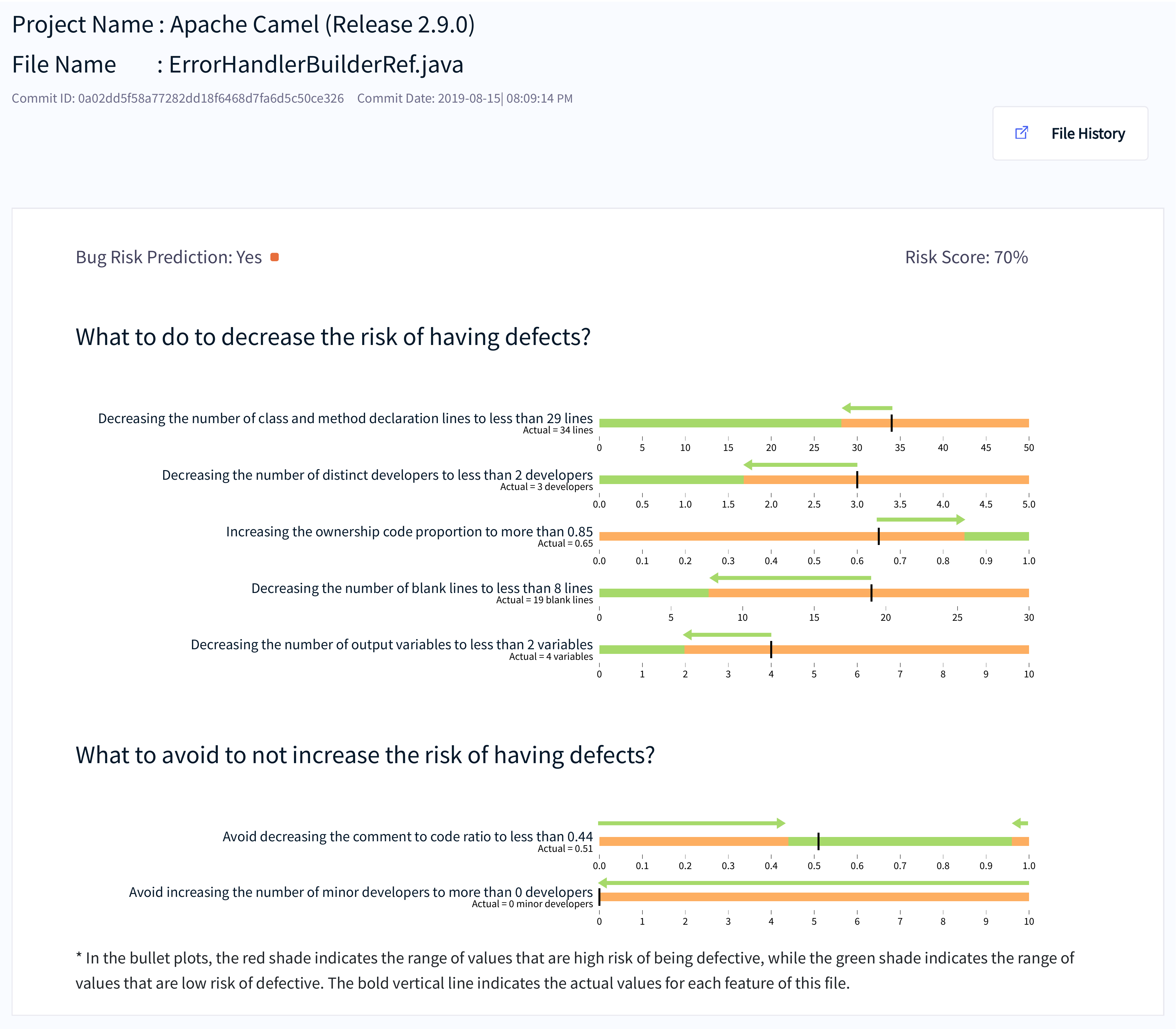}
\caption{The visualization of our SQAPlanner is designed to provide the following key information: (1) the list of guidance that practitioners should follow and should avoid; (2) the actual feature value of that file; and (3) its threshold and range values for practitioners to follow to mitigate the risk of having defects.}
\label{Fig:sqaplannerbulletplot}
\end{figure*}

\revisedTextOnly{
\section{Practitioners' Perceptions of our SQAPlanner Visualization}\label{sec:postvalidation}

In this section, we aim to investigate the practitioners' perceptions of 
the visualization of SQAPlanner when comparing to the visualization of the state-of-the-art (RQ6) and 
the actual guidance generated by SQAPlanner (RQ7).
Below, we describe the approach and present the results.


\revised{R1.2, R1.9, R2.13, R3.3}{\subsection{Approach}}

To address RQs 6 and 7, we developed a proof-of-concept to visualize the actionable guidance generated by our SQAPlanner. 
Traditionally, the importance scores of Random Forests or LIME's model-agnostic techniques are commonly presented using a bar chart. 
However, such bar charts can only indicate the importance scores, without providing guidance on what to do and what not to do.

To address this challenge, we propose to use a bullet plot (see Figure~\ref{Fig:sqaplannerbulletplot}).
The visualization of our SQAPlanner is designed to provide the following key information: (1) the list of guidance that practitioners should follow and should avoid; (2) the actual feature value of that file; and (3) its threshold and range values for practitioners to follow to mitigate the risk of having defects.
The green shades indicate the non-risky range values of features, while the red shades indicate the risky range values of features.
The vertical bars indicate the actual values of features for a given file.
The green arrows provide directions of how a feature should be changed (i.e., increase or decrease).
The list of guidance is structured into two parts: (1) what to do to decrease the risk of being defective; and (2) what to avoid to not increase the risk of being defective.
For each guidance, we translate a rule-based explanation into an actionable guidance.
A guidance is presented in the form of natural language to ensure that it is actionable and understandable by practitioners. 

To translate the rule-based explanations into actual guidance, we focus on only the \emph{ErrorHandlerBuilderRef.java} of the release 2.9.0 of the Camel software system.
We use the rule-based explanations from Section~\ref{sec:discussion} as a reference.
Finally, we derive the following statements according to the reference rule-based explanations in Section~\ref{sec:discussion}:

\begin{itemize}
    \item (S1) Decreasing the number of class and method declaration lines to less than 29 lines to decrease the risk of being defective.
    \item (S2) Decreasing the number of distinct developers to less than 2 developers to decrease the risk of being defective.
    \item (S3) Increasing the ownership code proportion to more than 0.85 to decrease the risk of being defective.
    \item (S4) Avoid decreasing the comment to code ratio to less than 0.44 to not increase the risk of being defective.
    \item (S5) Avoid increasing the number of minor developers to more than 0 developers to not increase the risk of being defective.
    \item (S6) Decreasing the number of blank lines to less than 8 lines to decrease the risky of being defective.
    \item (S7) Decreasing the number of output variables to less than 2 variables to decrease the risk of being defective.
\end{itemize}

To implement the visualization of our SQAPlanner approach, we decided to use the Microsoft's Code Defect AI as our core infrastructure.
We first downloaded the repository of Code Defect AI from GitHub.\footnote{https://github.com/aricent/codedefectai}
Then, we carefully studied their repository and deployed Code Defect AI in our local environment with continuous support from the core developer of Code Defect AI.
Then, we integrated our SQAPlanner approach and replaced their visualization (bar plots) with our visualization generated by SQAPlanner using the implementation of bullet plots as provided by the d3.js Javascript library.\footnote{\url{https://bl.ocks.org/mbostock/4061961}}

\begin{figure*}[t]
\centering
\begin{subfigure}[t]{.21\textwidth}
\includegraphics[width=.9\linewidth]{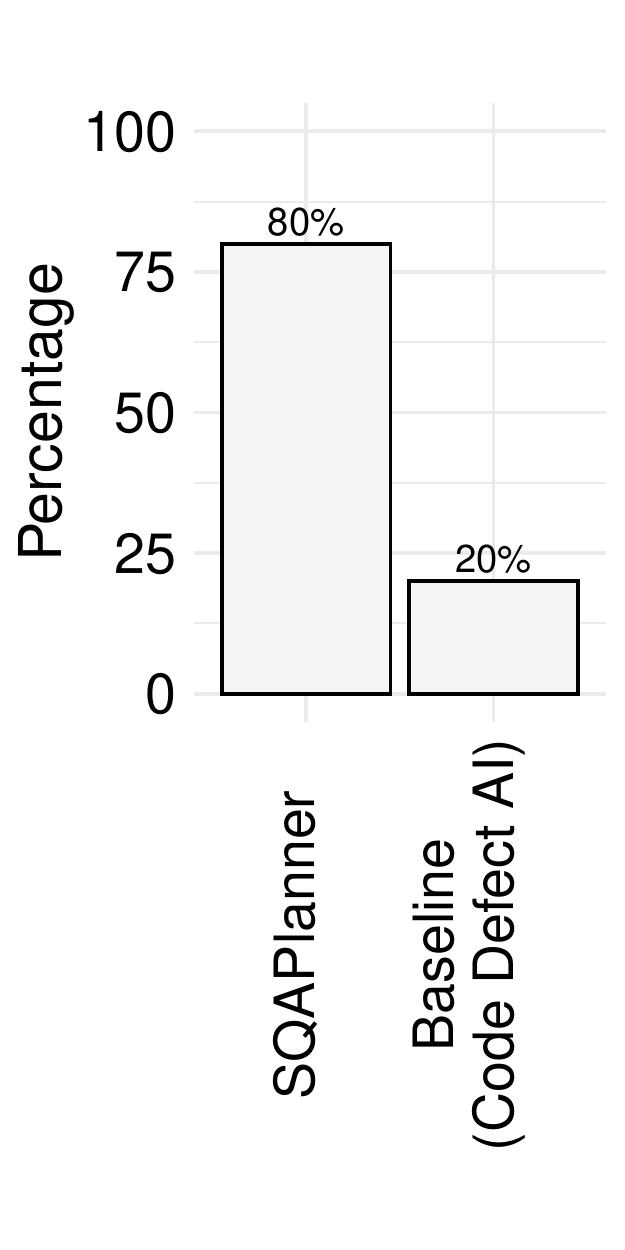}
\caption{Perceptions of visualization.}
\label{fig:rq6-results-1}
\end{subfigure} 
\begin{subfigure}[t]{.75\textwidth}
\includegraphics[width=.9\linewidth]{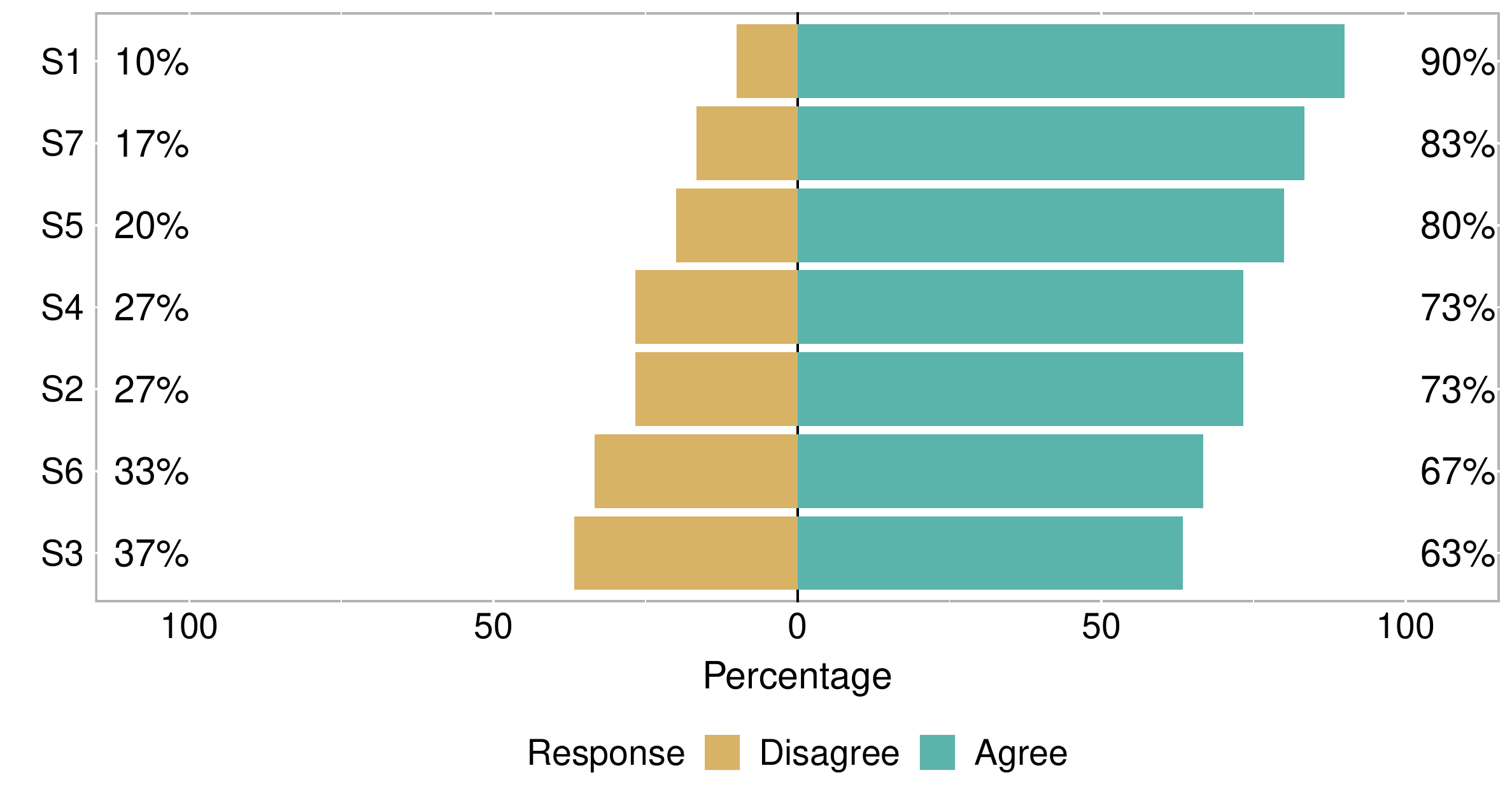}
\caption{Perceptions of the actual guidance generated by our SQAPlanner.}
\label{fig:rq7-results-1}
\end{subfigure} 
\caption{(RQ6,RQ7) The results of a qualitative survey with practitioners.}
\label{Fig:postsurveyresults}
\end{figure*}

To investigate the practitioners' perceptions of our SQAPlanner visualization, we used a qualitative survey as a research method.
We also used the visualization of Microsoft's Code Defect AI (see Figure~\ref{Fig:codedefectaiplot}) as a baseline comparison. 
The objectives of the survey are as follows: (1) to investigate the practitioners' perceptions of the visualization of our SQAPlanner; and (2) to investigate the practitioners' perceptions of the actionable guidance generated by our SQAPlanner.
Similar to Section~\ref{sec:survey}, we designed the survey as a cross-sectional study where participants provide their responses at one fixed point in time.
The design of our survey is described below.

\emph{Part 1---Practitioners' perceptions of the visualizations of our SQAPlanner:} 
We first provided the concept of defect prediction models and described how our SQAPlanner can be used to support SQA planning.
Then, we presented the visualization of our SQAPlanner and the visualization of Microsoft's Code Defect AI.
We asked the participants a closed-ended question to inquire about which of the visualization is the best to provide actionable guidance on how to mitigate the risk of having defects.
We also asked the participants an open-ended question to inquire about their rationale on why the selected visualization is preferred over another visualization.

\emph{Part 2---Practitioners' perceptions of the actual guidance generated by SQAPlanner:} 
We again presented the visualization of SQAPlanner.
Then, for each statement, we asked the participants a closed-ended question to inquire whether the participants agree for each of the seven statements that we translated from the rule-based explanations.

We used an online questionnaire service as provided by Google Forms.
We carefully evaluated the survey via pre-testing~\cite{litwin1995measure} to assess the reliability and validity of the survey.
The survey has been rigorously reviewed and approved by the Monash University Human Research Ethics Committee (MUHREC Project ID: 27209).
We used a recruiting service provided by the MTurk to recruit participants.
We received 240 closed-ended and 30 open-ended responses from 30 respondents.
Finally, we manually verified and analyzed the survey responses to ensure that the responses are of high quality.

\subsection{Results}
\subsection*{(RQ6) \rqvi}

\smallsection{Results}
\textbf{80\% of our respondents agree that the visualization of our SQAPlanner is better for providing actionable guidance when compared to the visualization of Microsoft's Code Defect AI.}
Figure~\ref{fig:rq6-results-1} shows the percentage of the respondents who select which visualization is best to provide actionable guidance on how to mitigate the risk of having defects.

After analyzing the open-end responses, practitioners (e.g., R10 and R12) provided rationales that the suggested threshold values of each factor and directional arrows provided by SQAPlanner make the visualization more clear on what developers should do and should avoid to decrease the risk of having defects.
Respondents (e.g., R19, R20, and R23) also pointed out that the summary of "What to do" and "What to avoid" is straight to the point and helpful.

On the other hand, 20\% of the respondents rate the visualization of Microsoft's Code Defect AI as better.
Respondents (e.g., R5 and R16) provided rationales that the visualization of Microsoft's Code Defect AI is presented in a more simple and concise manner (i.e., only present the most important factors that are associated with the risk of having defects).
Thus, future research should take into consideration the complexity of the provided information when designing a novel visualization for AI-driven defect prediction.

\subsection*{(RQ7) \rqvii}

\smallsection{Results}
\textbf{63\%-90\% of the respondents agree with the seven statements derived from the actual guidance generated by our SQAPlanner.}
Figure~\ref{fig:rq7-results-1} presents that the percentage of the respondents who agree with the seven statements derived from the actual guidance generated by our SQAPlanner.
We find that 90\% of the respondents agree the most with \emph{(S1) Decreasing the number of class and method declaration lines to less than 29 lines to decrease the risk of being defective}.

On the other hand, only 63\% of the respondents agree with \emph{(S3) Increasing the ownership code proportion to more than 0.85 to decrease the risk of being defective}.
We suspect that the wide range of agreement rates for our statements has to do with the degree of understandability of the software metrics, since practitioners may find that the number of class and method declaration lines for S1 is more intuitive and easy to understand than the ownership code proportion for S3.
Thus, future research should take into consideration the degree of understandability of the software metrics when designing a novel visualization for AI-driven defect prediction.}

\section{Threats to Validity}\label{sec:threats}

\smallsection{Construct Validity}
Many local model-agnostic techniques could be used to generate many forms of explanations e.g., feature importance and rules.
In this paper, we focused only on rule-based explanations by comparing with LORE and Anchor, an extension of LIME.
We also studied only a limited number of available  classification techniques. 
Thus, our results may not be applicable or generalise to the use of other techniques. 
Nonetheless, other classification techniques can be explored in future work to see if they improve on our results. 

There may be possible interpretation of causal relationship in our survey design.
Thus, we attempted to mitigate the negative effect caused by possible interpretation of causal relationship in the survey design as much as possible.
For example, we use the term "decrease/increase the risk of having defects", rather than "fixing the actual defects of that file".
We use the term "guidance" rather than "solutions", as these rule-based explanations are not the actual solutions to fix the actual defects of that file. 
Nevertheless, there may be other possible confounding factors that may cause interpretation of causal relationship. 
Thus, future studies should explore further.

In addition, we do not seek to claim the generalization and causation of the guidance generated by our approach. 
Instead, the key message of our study is that our rule-based guidance can explain the behaviour of the defect prediction models. 
Thus, they only indicate important relationships in the data and provide a useful tool to support decision- and policy-making in SQA planning activities.

\smallsection{Internal Validity}
The practicality of rule-based explanations heavily relies on software metrics that are used to train the models.
In this paper, we chose to generate rules based on 65 well-known and hand-crafted software metrics, rather than using advanced automated feature generation like deep learning.
Future work may focus on trying to explain other machine learning-based models, such as explaining deep learning models used in an SQA context.

\revised{R1.8}{The goal of our SQAPlanner (aka. the local rule-based model-agnostic technique) is a post-hoc analysis of the global defect prediction models. 
That means, SQAPlanner can only explain the behavior of the (global) defect prediction models, regardless of the correct or incorrect predictions. 
If the predictions of the global defect models for the testing dataset are incorrect, SQAPlanner will explain why the global defect prediction models generate wrong predictions in the form of rule-based explanations. 
Therefore, the robustness or the sensitivity of our SQAPlanner does not depend on the accuracy of the predictions of the global defect prediction models.}

\smallsection{External Validity}
We applied our SQA Planner approach to a limited number of software systems.
Thus, our results may not generalize to other datasets, domains, ecosystems.  
However, we mitigated this by choosing a range of different non-trivial, real-world, open-source software applications.
Nonetheless, additional replication studies in a proprietary setting and other ecosystems will prove useful to compare to our results reported here.

\revised{R2.11}{
SQA planning involves various activities.
However, this paper only focused on helping practitioners to define development policies and their associated risk thresholds~\cite{galin2018software}, without considering other activities.
In addition, the dependent variable that we used in this study only focused on software quality (i.e., defective or clean), without considering other aspects (e.g., testability, reusability, robustness, and maintainability).
Thus, other SQA planning activities and other quality attributes can be explored in future work.}

\section{Related Work}\label{sec:relatedwork}

\subsection{Explainable AI in Software Engineering}

Despite the advances of AI/ML techniques that are tightly integrated into software development practices (e.g., defect prediction~\cite{hall2012systematic}, automated code review~\cite{asthana2019whodo,thongtanunam2015should}, automated code completion~\cite{hellendoorn2019code,hellendoorn2018deep}), such AI/ML techniques come with their own limitations.
The central problem of AI/ML techniques is that most AI/ML models are considered black-box models i.e., we understand the underlying mathematical principles without explicit declarative knowledge representation.
In other words, developers do not understand how decisions are made by such AI/ML techniques.
In addition, the current defect modelling practices do not uphold the current data privacy laws and regulations, which require justifications of individual predictions for any decisions made by an AI/ML model.
Therefore, applying such black-box AI/ML techniques in the software development practices for safety-critical and cyber-physical systems~\cite{borg2019explainability,yang2017actionable} which involve safety, security, business, personal, or military operations is unfavourable and must be avoided.

Explainable AI is essential in software engineering to building appropriate trust (including Fairness, Accountability, and Transparency (FAT)).
Developers can then (1) understand the reasons and the logic behind every decision and (2) effectively improve the prediction models by understanding any unsound predictions made by the models.
Recently, explainable AI has been employed in software engineering~\cite{tantithamthavorn2020explainable,jiarpakdee2021perception}, by making defect prediction models more practical~\cite{pornprasit2021jitline,wattanakriengkrai2020predicting} (i.e., using LIME to explain which tokens and which lines are likely to be defective in the future) and explainable~\cite{jiarpakdee2020xai4se} (i.e, using LIME to explain a prediction why a file is predicted as defective).
However, there exists no studies that able to provide concrete guidance on what developers should do or should not do to support SQA planning.
To the best of our knowledge, this paper is the first to generate local rule-based explanations to help QA teams make data-informed decisions in software quality assurance planning.


\subsection{Towards Explainable and Actionable Analytics for Software Defects}

There are two key approaches for achieving explainability in defect prediction models.
The first is to make the entire decision process transparent and comprehensible (i.e., global explainability).
The second is to explicitly provide an explanation for each individual prediction (i.e., local explainability).


Examples of global explainability methods are regression models~\cite{rahman2013and,Rajbahadur2017}, decision trees~\cite{Zimmermann2009}, decision rules~\cite{rodriguez2012searching}, and Fast-and-Frugal trees \cite{chen2018applications}.
These transparent AI/ML techniques often provide built-in model interpretation techniques to uncover the relationships between the studied features and defect-proneness.
For example, an ANOVA analysis provided for logistic regression or a variable importance analysis provided for random forest.
However, the insights derived from these transparent AI/ML techniques do not provide justifications for each individual prediction.

Model-agnostic techniques are techniques for explicitly providing an instance explanation for each decision of AI/ML models (i.e., local explainability) for a given testing instance~\cite{Guidotti2018}.
Formally, given a defect model $f$ and an instance $x$, 
the instance explanation problem aims to provide an explanation $e$ for the prediction $f(x) = y$.
To do so, we address the problem by building a local interpretable model $f'$ that mimics the local behaviour of the global defect model $f$. 
An explanation of the prediction is then derived from the local interpretable model $f'$. 
The local interpretable model focuses on learning the behaviour of the defect models in the neighbourhood of the specific instance $x$, without aiming at providing a single description of the logic of the black box for all possible instances. 
Thus, an explanation $e \in E$ is obtained through $f'$, if $e = \epsilon(f', x)$ for some explanation logic $\epsilon(., .)$ which reasons over $f'$ and $x$.
Two common ways to represent explanations are feature-importance explanations and rule-based explanations.
Unlike model-specific explanation techniques discussed above, the great advantage of model-agnostic techniques is their flexibility. 
Such model-agnostic techniques can (1) interpret any learning algorithms (e.g., regression, random forest, and neural networks); (2) are not limited to a certain form of explanations (e.g., feature importance or rules); and (3) are able to process any input data (e.g., features, words, and images~\cite{ribeiro2016model}).

There are a plethora of model-agnostic techniques~\cite{Guidotti2018} for identifying the most important feature at the instance level.
For example, LIME (i.e., Local Interpretable Model-agnostic Explanations)~\cite{ribeiro2016should} is a model-agnostic technique that mimics the behaviour of the black-box model with a local linear model to generate the explanations of the predictions.
BreakDown~\cite{staniak2018explanations,gosiewska2019ibreakdown} is a model-agnostic technique that uses the greedy strategy to sequentially measure contributions of metrics towards the expected prediction.
However, none of these techniques can generate explanations with the logic behind.

Despite the advances of model-agnostic techniques in the Explainable AI communities, such techniques have not been employed in practical software engineering contexts.
To the best of our knowledge, this paper is the first to generate local rule-based explanations to help QA teams make data-informed decisions in software quality assurance planning.

\section{Conclusions}\label{sec:conclusions}

Defect prediction models have been proposed to generate insights (e.g., the most important factors that are associated with software quality).
However, such insights derived from traditional defect models are far from actionable---i.e., practitioners still do not know what they should do and should avoid to decrease the risk of having defects{, and what is a risk threshold for each metric.}
A lack of actionable guidance and its risk threshold could lead to inefficient and ineffective SQA planning processes.

In this paper, we investigate practitioners perceptions and their challenges of current SQA planning activities and the perceptions of our proposed four types of guidance.
Then, we propose and evaluate our SQAPlanner approach---i.e., an approach for generating four types of guidance and its risk threshold in the form of rule-based explanation for the predictions of defect prediction models.
Finally, we develop and evaluate the visualization of our SQAPlanner approach.

Through the use of qualitative survey and empirical evaluation, our results lead us to conclude that SQAPlanner is needed, important, effective, stable, and applicable.
We also find that 80\% of respondents perceived that our visualization is more actionable.
Thus, our SQAPlanner paves a way for novel research in actionable software analytics.

\textbf{Acknowledgments.} C. Tantithamthavorn was partially supported by the Australian Research Council's Discovery Early Career Researcher Award (ARC DECRA) funding scheme (DE200100941).
C. Bergmeir was partially supported by the Australian Research Council's Discovery Early Career Researcher Award (ARC DECRA) funding scheme (DE190100045).
J. Grundy was partially supported by the Australian Research Council's Laureate Fellowship funding scheme (FL190100035).


\bibliographystyle{IEEEtranS}
\bibliography{myref}

\begin{IEEEbiography}[{\includegraphics[width=1in,height=1.25in,clip,keepaspectratio]{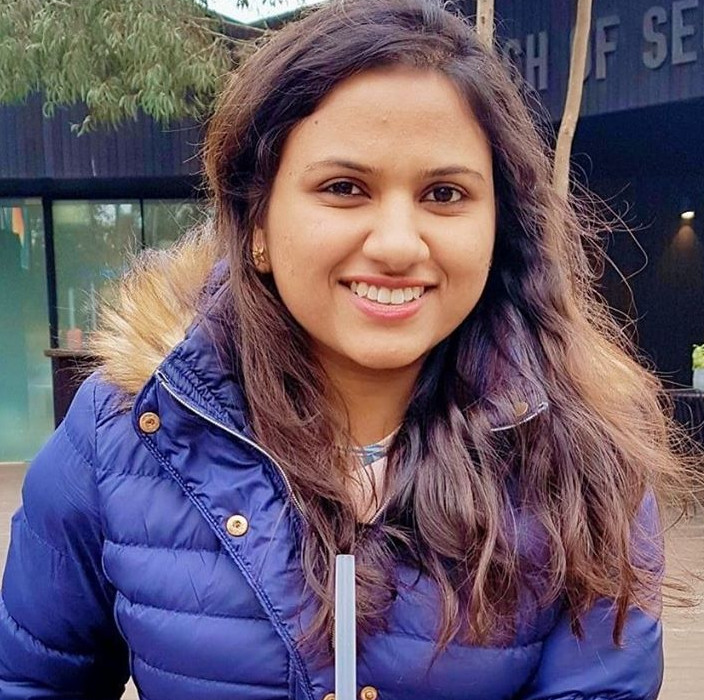}}]{Dilini Rajapaksha}
received the BSc(hons) degree from Sri Lanka Institute of Information Technology (SLIIT). 
She is currently a Ph.D. candidate at Monash University, Australia. 
Her research interests include Machine Learning and Time-series Forecasting. 
The goal of her Ph.D. is to provide local explanations for the predictions given by the time-series and machine learning models.
\end{IEEEbiography}

\begin{IEEEbiography}[{\includegraphics[width=1in,height=1.25in,clip,keepaspectratio]{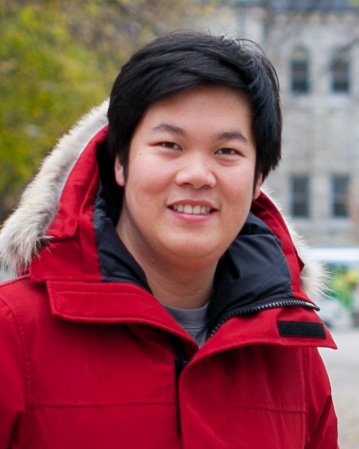}}]{Chakkrit Tantithamthavorn}
is a Lecturer in Software Engineering and a 2020 ARC DECRA Fellow in the Faculty of Information Technology, Monash University, Australia. 
His current fellowship is focusing on the development of ``Practical and Explainable Analytics to Prevent Future Software Defects''.
His work has been published at several top-tier software engineering venues, such as the IEEE Transactions on Software Engineering (TSE), the Springer Journal of Empirical Software Engineering (EMSE) and the International Conference on Software Engineering (ICSE). 
More about Chakkrit and his work is available online at \url{http://chakkrit.com}.
\end{IEEEbiography}

\begin{IEEEbiography}[{\includegraphics[width=1in,height=1.25in,clip,keepaspectratio]{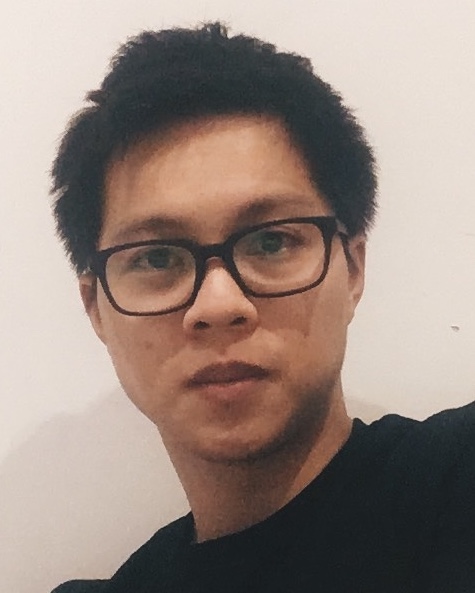}}]{Jirayus Jiarpakdee}
is a Ph.D. candidate at Monash University, Australia. His research interests include empirical software engineering and mining software repositories (MSR). The goal of his Ph.D. is to apply the knowledge of statistical modelling, experimental design, and software engineering to improve the explainability of defect prediction models.
\end{IEEEbiography}

\begin{IEEEbiography}[{\includegraphics[width=1in,height=1.25in,clip,keepaspectratio]{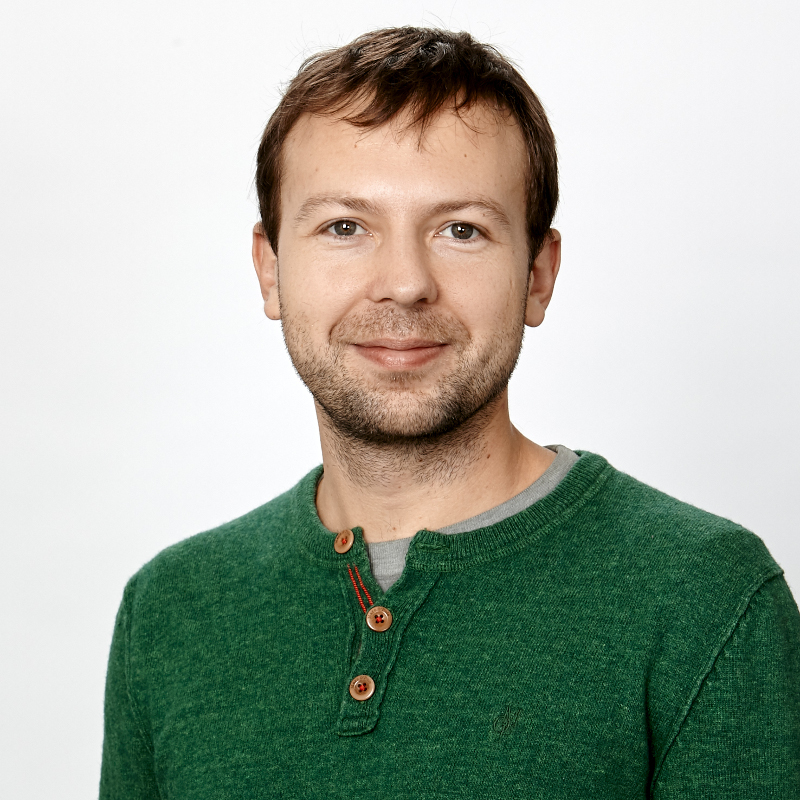}}]{Christoph Bergmier}
is a Lecturer in Data Science and Artificial Intelligence, and a 2019 ARC DECRA Fellow in the Monash Faculty of Information Technology. His fellowship is on the development of "efficient and effective analytics for real-world time series forecasting". He also works as a Data Scientist in a variety of projects with external partners in diverse sectors, e.g. in healthcare or infrastructure maintenance. Christoph holds a PhD in Computer Science from the University of Granada, Spain, and an M.Sc. degree in Computer Science from the University of Ulm, Germany.
\end{IEEEbiography}

\begin{IEEEbiography}[{\includegraphics[width=1in,height=1.25in,clip,keepaspectratio]{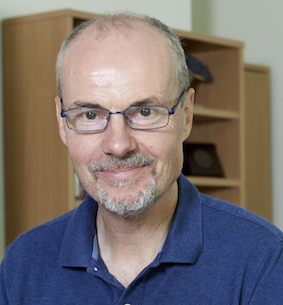}}]{John Grundy} is Australian Laureate Fellow and Professor of Software Engineering at Monash University, Australia. He has published widely in automated software engineering, domain-specific visual languages, model-driven engineering, software architecture, and empirical software engineering, among many other areas. He is Fellow of Automated Software Engineering and Fellow of Engineers Australia.
\end{IEEEbiography}

\begin{IEEEbiography}[{\includegraphics[width=1in,height=1.25in,clip,keepaspectratio]{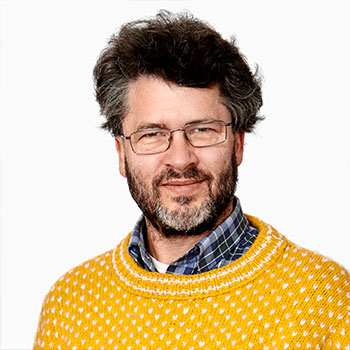}}]{Wray Buntine}
is a Professor in Machine Learning in the Faculty of Information Technology at Monash University in Melbourne, Australia, and a recipient of the 2019 Google AI Impact Challenge.
He is known for his theoretical and applied work, probabilistic methods for document and text analysis, social networks, data mining and machine learning. 
His work is supported by many high-profile organisations such as Google and the NASA Ames Research Centre, as well as Wall Street and Silicon Valley startups.
\end{IEEEbiography}
\end{document}